\documentclass[journal,a4paper]{IEEEtran}

\usepackage[acronyms, nonumberlist, nopostdot, nomain, nogroupskip]{glossaries}
\usepackage{amsmath,amsfonts,amssymb,amsthm}
\usepackage{bm}
\usepackage{mathtools}
\usepackage{algorithm,algpseudocode}
\usepackage{array}
\usepackage[caption=false]{subfig}
\usepackage{url}
\usepackage{verbatim}
\usepackage{graphicx}
\usepackage{cite}
\usepackage{enumerate}
\usepackage[normalem]{ulem}
\usepackage{tikz}
\usetikzlibrary{shapes, arrows, positioning, plotmarks, patterns}    
\usepackage{pgfplots}
\pgfplotsset{
compat=1.18,
legend style={font=\footnotesize, fill opacity=0.7,  draw opacity=1, text opacity=1, draw=white!15!black, legend cell align=left, align=left}, 
yminorticks=false,
xminorticks=false,
title style={font=\small},
tick style={color=black},
every tick label/.append style={font=\footnotesize},
tick label style={font=\small},
grid style={line width=.1pt, draw=gray!20},
major grid style={line width=.1pt,draw=gray!20},
}
\usepgfplotslibrary{colorbrewer, groupplots, fillbetween}

\newcommand{\T}{^{\intercal}}     
\newcommand{\E}[1]{\mathbb{E}\bigg[ #1 \bigg]} 
\newcommand{\mc}[1]{\mathcal{#1}}   
\newcommand{\mb}[1]{\mathbf{#1}}    
\DeclareMathOperator*{\argmax}{arg\,max}    
\DeclareMathOperator*{\diag}{D}    
\newcommand{\set}[1]{\mathcal{#1}}

\algtext*{EndWhile}
\algtext*{EndIf}
\algtext*{EndFor}

\newtheorem{theorem}{Theorem}

\newtheorem{lemma}{Lemma}[theorem]

\newtheorem{corollary}{Corollary}[theorem]

\def \sfwidth{0.24\linewidth}
\def \sfheight{0.20\linewidth}

\def \hh{0.475\linewidth}
\def \ww{0.99\linewidth}

\definecolor{lightgray}{HTML}{999999}
\definecolor{color1}{HTML}{440154}
\definecolor{color2}{HTML}{31688e}
\definecolor{color3}{HTML}{35b779}
\definecolor{color4}{HTML}{ffa600}

\newacronym{rl}{RL}{reinforcement learning}
\newacronym{marl}{MARL}{multi-agent reinforcement learning}
\newacronym{mdp}{MDP}{Markov decision process}
\newacronym{pomdp}{POMDP}{partially observable Markov decision process}
\newacronym{acnomdp}{ACNO-MDP}{action-contingent noiselessly observable MDP}
\newacronym{aoi}{AoI}{age of information}
\newacronym{uoi}{UoI}{urgency of information}
\newacronym{voi}{VoI}{value of information}
\newacronym{ncs}{NCS}{networked control system}
\newacronym{pi}{PI}{policy iteration}
\newacronym{api}{API}{alternating policy iteration}
\newacronym{mpi}{MPI}{modified policy iteration}
\newacronym{ne}{NE}{Nash equilibrium}
\newacronym{ibr}{IBR}{iterated best response}
\newacronym{decpomdp}{Dec-POMDP}{decentralized partially observable Markov decision process}
\newacronym{fip}{FIP}{finite improvement property}
\newacronym{cps}{CPS}{cyber-physical system}
\newacronym{sarsop}{SARSOP}{successive approximations of the reachable space under optimal policies}
\newacronym{pls}{PLS}{polynomial local search}
\newacronym{jpo}{JPO}{joint policy optimization}

\usepackage{layout}
\usepackage[top=1.9cm,bottom=4.3cm,left=1.35cm,right=1.36cm]{geometry}

\begin{document}

\setlength{\textheight}{665pt}
\setlength{\columnsep}{12pt}
\title{Pragmatic Communication for Remote Control\\of Finite-State Markov Processes}

\author{Pietro Talli, Edoardo David Santi, Federico Chiariotti, Touraj Soleymani,\\ Federico Mason, Andrea Zanella,  and Deniz G\"{u}nd\"{u}z%
\thanks{Part of this work was funded by the European Union under the Italian National Recovery and Resilience Plan (NRRP) of NextGenerationEU, underpartnership on “Telecommunications of the Future” (PE0000001 - Program “RESTART”). Limited partial versions of this work were previously published\,/\,submitted as two conference papers~\cite{talli2024push,santi2024remote}. Talli and Edoardo David Santi contributed equally to this work. Federico Chiariotti and Touraj Soleymani also contributed equally to this work. Pietro Talli, Federico Mason, Federico Chiariotti, and Andrea Zanella are with the Department of Information Engineering, University of Padova, 35131 Padua, Italy (emails: {\scriptsize\tt \{tallipietr, masonfed, chiariot, zanella\}@dei.unipd.it}). Edoardo David Santi, Touraj Soleymani, and Deniz G\"{u}nd\"{u}z are with the Department of Electrical and Electronic Engineering, Imperial College London, London SW7 2AZ, United Kingdom (e-mails: {\tt\scriptsize \{edoardo.santi17, touraj, d.gunduz\}@imperial.ac.uk}). Touraj Soleymani is also with the City St George’s School of Science and Technology, University of London, London EC1V 0HB, United Kingdom.}
\vspace{-0.5cm}
}
\maketitle

\begin{abstract}
Pragmatic or goal-oriented communication can optimize communication decisions beyond the reliable transmission of data, instead aiming at directly affecting application performance with the minimum channel utilization. In this paper, we develop a general theoretical framework for the remote control of finite-state Markov processes, using pragmatic communication over a costly zero-delay communication channel. To that end, we model a cyber-physical system composed of an encoder, which observes and transmits the states of a process in real-time, and a decoder, which receives that information and controls the behavior of the process. The encoder and the decoder should cooperatively optimize the trade-off between the control performance (i.e., reward) and the communication cost (i.e., channel use). This scenario underscores a pragmatic (i.e., goal-oriented) communication problem, where the purpose is to convey only the data that is most valuable for the underlying task, taking into account the state of the decoder (hence, the pragmatic aspect). We investigate two different decision-making architectures: in pull-based remote control, the decoder is the only decision-maker, while in push-based remote control, the encoder and the decoder constitute two independent decision-makers, leading to a multi-agent scenario.
We propose three algorithms to optimize our system (i.e., design the encoder and the decoder policies), discuss the optimality guarantees ofs the algorithms, and shed light on their computational complexity and fundamental limits.
\end{abstract}

\begin{IEEEkeywords}
Effective communication, goal-oriented communication, cyber-physical systems, implicit information, Markov decision processes.
\end{IEEEkeywords}

\vspace{-3mm}
\glsresetall
\section{Introduction}
\label{sec:intro}

\IEEEPARstart{T}{he paradigm} of \glspl{cps} has recently enabled the creation of complex system architectures that tightly integrate communication, computation, and control \cite{baillieul2007, kim_cyberphysical_2012}. \Glspl{cps} often exhibit a dynamic and distributed nature~\cite{lee2023}, which necessitates persistent status updating of local components to reflect changes in the environment~\cite{popovski2024time}. This steady influx of real-time data empowers a \gls{cps} to swiftly adapt to evolving conditions, ensuring that control decisions are made based on the most pertinent and up-to-date information~\cite{touraj-power}.
The performance of a \gls{cps} is, therefore, dramatically affected by both control and communication policies~\cite{busoniu2008comprehensive}. However, modeling, design, and optimization of \Glspl{cps} in terms of control and communication policies can be quite challenging when control quality indices and communication constraints are simultaneously taken into account~\cite{gunduz2022semantic, uysal2022semantic}.

Prior research has attempted to address these challenges by optimizing the \gls{aoi}, a metric that captures the freshness of information in real-time networked systems~\cite{yates2021age}. Unfortunately, policies based on the \gls{aoi} may result in abrupt changes in the process states being unreported for a relatively long time, as well as wasting resources for communication of irrelevant data~\cite{banerjee2020fundamental}. This has led to the development of the notion of the \gls{voi} in the context of \glspl{cps}~\cite{soleymani_value_2022, soleymani_value_2023}, which captures the significance of information and is intimately related to pragmatic communication\footnote{In the literature, pragmatic communication is also known as goal-oriented, task-oriented, or effective communication.} approaches~\cite{gunduz2023timely}. The recent developments in communications focus on approaches that integrate semantic and pragmatic aspects~\cite{uysal2022semantic} into the communications pipeline~\cite{maatouk2022age}, which go beyond Shannon's strict source-channel separation and tailor communication decisions such as data compression and power scheduling to specific application goals~\cite{touraj-power,talli2024effective}. Nevertheless, the interactions between control and communication policies are non-trivial in complex scenarios, requiring further developments for solving intractable optimization problems in a cooperative~way~\cite{goldman2004decentralized}.

In the present work, we consider a finite-state Markovian physical process to be remotely monitored and, possibly, controlled, over a costly zero-delay communication channel. We model the system as a two-agent \gls{decpomdp}, in which the first agent, named \textit{encoder}, observes and transmits the states of the process in real time, and the second agent, named \textit{decoder}, receives that information and controls the behavior of the process. In this system, on one hand, what the decoder receives depends on the communication decisions regarding transmitting the status updates over the channel. On the other hand, what the encoder observes in the future depends on the control decisions that can affect the process' state evolution. We investigate two different decision-making architectures. In the first one, called \textit{pull-based} architecture, the decoder is the only decision-maker and is in charge of both communication and control decisions. In the second, called \textit{push-based} architecture, the encoder and the decoder constitute two independent decision-makers, responsible for communication and control decisions, respectively. Intuitively, the former is simpler, while the latter can lead to higher performance due to an extended information structure.

Although the literature includes multiple solutions for the remote estimation and control of Markov processes, a unified theory for handling these tasks is still missing.
Many works investigate heuristic or data-driven algorithms that, despite providing high performance under specific conditions, are not general enough and do not take into account the game theoretical limitations of multi-agent systems.
In this work, we aim to fill the gaps in the literature, by designing a new and more general framework for the remote control of Markov processes,
that can effectively represent several real-world problems and provide a theoretical grounding to pragmatic communication schemes.
By doing so, we want to highlight and analytically model the subtle optimization challenges that characterize \glspl{cps}, especially in the case of limited bandwidth or other communication resources.
One of these challenges is associated with the exploitation of \emph{implicit information}, i.e., knowledge about the state of the process that the decoder can infer when the encoder does not transmit any data.


This paper builds upon our preliminary findings reported in~\cite{talli2024push, santi2024remote}:
in particular, we analyze the key features of pull-based and push-based architectures in~\cite{talli2024push}, while we investigate the benefits of implicit information in the push-based architecture in~\cite{santi2024remote}. The current work significantly improve these results by integrating our techniques in a unified model, highlighting the challenges associated with the joint design of control and communication policies, and formally proving the correctness and the limitations of the proposed schemes. This generalization extends to broader theoretical and numerical results, as well as computational complexity considerations.
Our ultimate goal is to develop a theoretical framework for solving the problem of the remote control of finite-state Markov processes over costly zero-delay communication channels either exactly or approximately, and delineating the fundamental properties of possible solutions.
In short, the main contributions of this paper are~as~follows:
\begin{enumerate}[\itshape(i)]
    \item We introduce a \gls{cps} model that can effectively represent pragmatic communication in remote control tasks, and can readily be specialized to remote estimation tasks. This model, in general, consists of an encoder and a decoder, which can exchange information and should cooperatively accomplish a goal. Based on this model, we investigate two decision-making architectures, namely, pull-based and push-based;
    \item In the pull-based setting, we propose the \gls{mpi} scheme, proving that it can reach an $\varepsilon$-optimal solution in polynomial time. However, we also prove that the pull-based setting represents a restricted version of the problem, with inherent performance limits;
    \item  In the push-based setting, we propose two schemes, namely, \gls{api}, which converges to a locally $\varepsilon$-optimal solution in polynomial time, and \gls{jpo}, which can achieve a globally $\varepsilon$-optimal solution but requires exponential computation time.
    We discuss formally the computational complexity and the fundamental limits of these schemes;
    \item We validate our theoretical results by means of experiments for remote estimation and control tasks, comparing the performance of the proposed schemes in a simulated environment. Our results, in particular, show that practical polynomial-time algorithms seem to be robust in the remote control tasks, while the optimality gap becomes much wider in the remote estimation tasks.
\end{enumerate}

The rest of the paper is organized as follows. We discuss the related work on remote control and remote estimation as well as on pragmatic communication in Sec.~\ref{sec:related}. Then, we introduce the system model and formulate the problem of interest in Sec.~\ref{sec:system}. We propose our algorithms in Sec.~\ref{sec:solution}, and provide our theoretical results in Sec.~\ref{sec:theory}. Then, we present our numerical analysis and simulation results in Sec.~\ref{sec:results}. Finally, we provide our concluding remarks in Sec.~\ref{sec:conc}. 

\section{Related Work}\label{sec:related}
In feedback control, the effectiveness of control actions directly depends on the quality of state estimates. This implies that, in order to tackle a remote control problem, one should be able to address first the corresponding remote estimation problem, as the most foundational task. The design of optimal policies for remote estimation and control of dynamical processes over communication channels has frequently been addressed in the literature.
Notably, the work in~\cite{walrand_optimal_causal_coding_1983} analyzes the best strategies for lossy transmissions without any communication cost in a finite time horizon.
The optimality of these strategies in the case of an infinite time horizon is proven in~\cite{wood_optimal-zero-delay-coding}, along with the $\varepsilon$-optimality of finite-memory quantizers.
The same work shows that deterministic policies lead to optimal encoding solutions, while the work in~\cite{cregg_reinforcement-learning-for-zero-delay-coding} adopts a \gls{rl} approach to address the target problem.
These works generally deal with lossy encoding, i.e., with the decision over how to encode system states with a fixed packet length.
The general properties of encoding policies for remote estimation are discussed in~\cite{chakravorty_fundamental_2016}, which shows that the trade-off curve between the estimation error and the transmission rate can be modeled by a piece-wise linear convex decreasing function.
However, these results only hold under specific conditions, i.e., for Markov processes with a rigid structure that simplifies the problem.

The problem of the remote estimation of scalar continuous-state Markov processes was addressed in~\cite{lipsa_remote_2011, molin2017}, where the optimal policies are also derived.
A more recent study in~\cite{erasure2023} investigates the same problem
in the presence of delay and packet loss, shedding light on the structure of the optimal policies and the role of implicit information.
However, the findings of these past works
cannot be generalized to finite-state Markov processes.
In~\cite{salimnejad_state-aware_2023}, the authors introduce the channel noise and design an encoding policy for monitoring a two-state Markov chain, taking into account the importance of each state in terms of the actions to be taken by the agent.
The same framework is extended to $N$-state Markov chains in \cite{salimnejad_real-time_2023}, which investigates the optimal configuration for a sampling and communication policy to minimize the reconstruction error under different constraints.
Other works~\cite{krale_act-then-measure_2023-2, bellinger_active-measure-RL_2021, nam2021reinforcement} propose solutions for single-agent Markov processes in which an agent needs to pay a fixed price to either observe the current or the next state.
This model suits pull-based settings, where a receiver must request state updates from a sensor over a resource-constrained channel.

Finally, several works have proposed learning-based pragmatic communication solutions for remote control~\cite{foerster2016learning}, showing that a significant performance improvement is obtained by the joint training of sensors and agents (i.e., encoders and decoders)~\cite{TungJSAC, mason2024multi}.
These works usually exploit the \gls{marl} paradigm, modeling the channel as an information bottleneck~\cite{wang2020learning} and considering the mutual information between agent beliefs~\cite{kim2023variational} as part of the system state.
However, \gls{marl} solutions provide no theoretical guarantees hard to adapt to scenarios different from those in which they were initially trained.

Indeed, there is a significant gap between the theoretical results on remote estimation and learning-based pragmatic communication schemes: while the former mostly give the optimal encoding in specific scenarios, the latter are often presented without any optimality guarantees, limiting their generality and trustworthiness.
Our work contributes to filling this gap by providing formal theoretical results for the general case of finite-state Markov processes with two agents, which can exchange data across a costly communication channel with the objective of either monitoring or controlling the process.
With respect to the state of the art, our study unifies recent results about the performance of pull-based and push-based pragmatic communication approaches, extending previous analyses by deriving computational complexity and optimality bounds.


\section{Problem Formulation and Modeling}\label{sec:system}
Let us consider a \gls{cps} model with two agents: an \emph{encoder} and a \emph{decoder}.
At each time step, the encoder observes the state of a discrete-time Markov process and should transmit this information to the decoder, whose task is to control the behavior of the process.
Our objective is to design the encoder's and the decoder's strategies in order to optimize the trade-off between the control performance (i.e., reward) and the communication cost (i.e., channel use) over an infinite time horizon.
This problem can be expressed mathematically as a two-agent \gls{decpomdp}~\cite{bernstein2002complexity} characterized by the tuple $\set{M}=\langle \set{S}, \set{O}, \set{A}, \set{C}, \mb{P}, o, r, \gamma \rangle$\footnote{In this work, we consider an infinite-horizon discounted formulation, but the theoretical and practical considerations can be adapted to related problems which aim at maximizing the average reward over a finite or infinite horizon.},~where
\begin{itemize}
    \item $\set{S}$ is the discrete and finite set of states of the underlying Markov chain; the state at time $t$ is denoted by $s_t$;
    \item $\set{O}=\mathcal{S} \,\cup \{\chi\}$ is the discrete and finite set of possible observations of the decoder, where the symbol $\chi$ represents the absence of transmission; the observation of the decoder at time $t$ is denoted by $o_t$. Note that the observation set of the encoder is $\set{S}$, as the encoder can directly observe the system state at each time;
    \item $\set{A}$ is the discrete and finite set of possible actions of the decoder; the control action at time $t$ is denoted by $a_t$;
    \item $\set{C}=\{0,1\}$ is the set of possible encoder actions (actions $0$ and $1$ correspond to no transmission and transmission of the current state, respectively), and the communication action at time $t$ is denoted by $c_t$;
    \item $\mb{P}\!\in\![0,1]^{|\set{S}|\!\times\!|\set{A}|\!\times\!|\set{S}|}$ is the transition matrix, whose entry $(s,a,s')$ represents the probability of moving to state $s'$ when performing action $a \in\set{A}$ in state $s$;
    \item $o:\set{S}\times\set{C} \mapsto\set{O}$ is the decoder observation function\footnote{In the \gls{decpomdp} literature, this commonly depends on the last control action $a$ as well. As $a$ is irrelevant in our model, we drop it for brevity's sake.}
    \begin{equation}
        o_t=o(s_t,c_t)=\begin{cases}
        s_t, & \text{if } c_t=1, \\
        \chi, & \text{if } c_t = 0;
    \end{cases}
    \end{equation}
    \item $r:\mc{S} \times \mc{A} \times \mc{S} \mapsto \mathbb{R}$ is the reward function, and the reward at time $t$ is given by $r_t = r_{s_t,s_{t+1}}(a_t)$;
    \item $\gamma\in[0,1)$ is the exponential discount factor.
\end{itemize}
At each time step, if the encoder transmits a status update, the decoder will have perfect knowledge about the current state of the process. Otherwise, the decoder will have imperfect knowledge, and must rely on its own history of past observations and actions to control the system.

We represent encoding and decoding policies by $\pi_{\text{e},t}$ and $\pi_{\text{d},t}$, respectively. These policies are causal, meaning that at each time step they depend only on the knowledge of the encoder and the decoder at that time step. To jointly optimize performance under a communication constraint, one needs to solve the following problem:
\begin{align}\label{eq:constrained_problem}
\text{maximize } \ \mathbb{E} \bigg[ \sum_{t=0}^\infty \gamma^t r_t \bigg], \text{ subject to } \ \mathbb{E} \bigg[ \sum_{t=0}^\infty \gamma^t c_t \bigg] \leq C,
\end{align}
where $C$ is value specifying a constraint on the cumulative channel use.
The above problem is a constrained \gls{decpomdp}, which may require stochastic policies. In this work, we reformulate \eqref{eq:constrained_problem} as an unconstrained \gls{decpomdp}, allowing us to safely restrict our attention to deterministic policies~\cite{Oliehoek2008}:
\begin{align}
\label{eq:unconstrained_problem}
    \text{maximize } \ \mathbb{E}\bigg[ \sum_{t = 0}^\infty \gamma^t (r_t - \beta c_t) \bigg],
\end{align}
where $\beta\in\mathbb{R}^+$ is a weighting coefficient. Note that $\beta$ is in fact the price that the system has to pay for the \emph{reliable communication} of a message from the encoder to the decoder.

Our first result is the following theorem, which dramatically simplifies the structures of the encoding and decoding policies.

\begin{theorem}\label{th:belief}
Without any loss of optimality, at each time $t$, the knowledge of the encoder can be described by $\langle s_t,\Delta_t,s_{t-\Delta_t}\rangle$, and that of the decoder by $\langle \Delta_t,s_{t-\Delta_t}\rangle$, where $\Delta_t$ is the time elapsed since the last transmission.
\end{theorem}
\begin{IEEEproof}
Note that the state of a \gls{mdp} is a sufficient statistic for the decision made by its underlying agent. In our problem, for any fixed decoder policy, at each time $t$, the encoder experiences a Markovian environment whose state is $\langle s_t, o_{0:t-1} \rangle$. Since this process is Markovian, the state $\langle s_t, o_{0:t-1} \rangle$ embodies all the information from the past evolution of the process. Moreover, the observation $o_{t}$ at the decoder depends on the encoder action $c_t$ that, in turn, only depends on $\langle s_t, o_{0:t-1} \rangle$. In addition, the current belief of the decoder is only a function of its history of observations and control actions since the last renewal instant, i.e., the last successful transmission. Note that the encoder provides either no explicit information to the decoder when $c_t=0$, or the exact state $s_t$ when $c_t=1$. Therefore, we deduce that the tuple $\langle \Delta_t,s_{t-\Delta_t}\rangle$ is a sufficient statistic for the decoder. Finally, following the same logic, we also deduce that the tuple $\langle s_t,\Delta_t,s_{t-\Delta_t}\rangle$ is a sufficient statistic for the~encoder.
\end{IEEEproof}

By Theorem~\ref{th:belief}, the decoder can then use $\langle \Delta_t,s_{t-\Delta_t}\rangle$ as a Markovian state. Hence, the decoder selects an action $a_t\in\mc{A}$ according to a control policy $\bm{\pi}_{\text{d}}:\set{S}\times\{0,\ldots,T_{\max}\}\mapsto\mc{A}$ (and also the following communication action $c_{t+1}\in\mc{C}$ in a pull-based system), gaining an instantaneous reward of $r_t = r_{s_t,s_{t+1}}(a_t)$, whose long-term maximization represents the system goal. 
To maintain a finite state space, we assume that the system allows a maximum inter-transmission time of $T_{\max}$, after which a new communication is always triggered.
In particular, we can set $T_{\max}$ to a value large enough to make the boundary approximation error arbitrarily small. We can also avoid considering a boundary when shifting to a belief-based problem formulation, as we will discuss later.

There can be two possible decision-making architectures for the optimization problem in (\ref{eq:unconstrained_problem}): \emph{pull-based} and \emph{push-based} architectures. In the pull-based setting, the decoder is the only decision maker and has to decide whether to request status updates and to select control actions, reducing the system to a single-agent problem. However, in the push-based setting, the encoder and the decoder are two independent decision-makers and must adapt to each other. Both settings have several practical applications, and the choice between the two also depends on architectural, energy, and cost considerations.

We highlight that our results can be easily extended to more realistic communication channels with time delay and a fixed packet loss rate, as long as the information structure is preserved, i.e., in these cases, the delay should be known to both nodes so that the encoder can compute the decoder's belief; an acknowledgment mechanism should also be implemented, so that the encoder knows the outcome of its transmissions.

\subsection{Pull-Based Remote Control}
\label{sub:pull}
The pull-based setting describes a restricted single-agent problem in which the system is entirely controlled by the decoder, and can be mapped to a problem class known as \gls{acnomdp}~\cite{nam2021reinforcement}, in which the state $s_t$ is available to the agent through a (costly) sensing action corresponding to a communication request to the encoder, which has perfect state information and can transmit it on demand. If we fix the communication policy, the resulting system is a standard POMDP and can be solved near-optimally~\cite{pineau2003point,smith2005point}.

Practically, each time step can be organized into two substeps, i.e., a communication step followed by a control step. During the first substep, the decoder must decide whether to request a new update from the transmitter, according to a policy $\bm{\tau}: \set{S}\times\{0,\ldots,T_{\max}\} \mapsto \mc{C}$. Following Theorem~\ref{th:belief}, this is equivalent to computing the full \emph{a priori} belief distribution given the history of the last $\Delta_t$ actions $\mb{a}_{t-\Delta_t:t-1}=(a_{t-\Delta_t}, a_{t-\Delta_t+1}, \ldots, a_{t-1})$, i.e.,
\begin{align}
 \bm{\omega}_{\langle \Delta_t,s_{t-\Delta_t}\rangle}^{\mb{a}_{t-\Delta_t:t-1}}=\bigg(\prod_{\ell=t-\Delta_t}^{t}\mb{P}^{a_{\ell}}\bigg)\T\mb{u}_{s_{t-\Delta_t}},\label{eq:belief}
\end{align}
where $\mb{u}_s$ is a one-hot column vector whose elements are all 0, except for the one corresponding to $s$, which is equal to 1, $\mb{P}^a$ is the transition matrix when taking action $a$, and $(.)\T$ denotes the transpose operation.
As the transmission process relies only on the decoder's estimations of the system state, the \emph{a posteriori} belief distribution is identical to the \emph{a priori} one. During the second substep, the decoder selects a control action according to policy $\bm{\pi}_{\text{d}}: \set{S}\times\{0,\ldots,T_{\max}\} \mapsto \mc{A}$. Between the two substeps, the decoder's state will change from $\langle\Delta_t,s_{t-\Delta_t}\rangle$ to $\langle 0,s_t\rangle$ if an update is requested, i.e., $c_t=1$, resetting the belief to a singleton, while $\bm{\omega}_t$ will be updated in a Bayesian manner using the transition matrix~$\mb{P}$ otherwise.

\subsection{Push-Based Remote Control}
The push-based setting corresponds to a two-agent problem, in which the encoder acts first, and the knowledge of the decoder is affected by the encoder's update.
Practically, at each time step $t$, the encoder uses policy $\bm{\pi}_{\text{e}}$ to decide whether to send a status update, while the decoder only determines control actions. Therefore, the communication action is determined by the policy $\bm{\pi}_{\text{e}}:\mc{S}\times\{0,T_{\max}\}\times\mc{S}\mapsto\mc{C}$. This two-agent problem is a fully cooperative Markov game with asymmetric information, as the encoder knows everything about the system, while the decoder does not. It is also an exact potential game~\cite{monderer1996potential}, as the reward of the two agents is exactly~the~same.

As the channel is zero-delay and lossless, the decoder knows that the encoder has decided not to transmit in the last $\Delta_t$ states $s_{t-\Delta_t+1},\ldots,s_t$. Accordingly, the \emph{a posteriori} belief distribution can be expressed recursively for $\Delta_t \geq 1$~as:
\begin{equation}\label{eq:belief_full}
\bm{\omega}^{\mb{a}_{t-\Delta_t:t-1},\bm{\pi}_{\text{e}}}_{\langle \Delta_t,s_{t-\Delta_t}\rangle}\!=\!\frac{\diag(\mb{1}\!-\!\mb{c}( \Delta_t,s_{t-\Delta_t}))\!\left(\mb{P}^{a_{t-1}}\right)\T\!\bm{\omega}^{\mb{a}_{t-\Delta_t:t-2},\bm{\pi}_{\text{e}}}_{\langle \Delta_t-1,s_{t-\Delta_t}\rangle}}{(\mb{1}-\mb{c}( \Delta_t,s_{t-\Delta_t}))\left(\mb{P}^{a_{t-1}}\right)\T\!\bm{\omega}^{\mb{a}_{t-\Delta_t:t-2},\bm{\pi}_{\text{e}}}_{\langle \Delta_t-1,s_{t-\Delta_t}\rangle}},
\end{equation}
where function $\diag(\mb{x})$ returns a diagonal matrix whose $i$-th diagonal entry is the $i$-th element of vector $\mb{x}$, and $\mb{c}( \Delta_t,s_{t-\Delta_t})$ is a row vector of length $|\mc{S}|$ whose $s$-th entry is $\pi_{\text{e}}( s,\Delta_t,s_{t-\Delta_t})$.

While the belief update in~\eqref{eq:belief} is naive, as it does not consider the encoder's policy, the push-based scenario presents higher complexity because of the existence of \emph{implicit information}.
By not transmitting, the encoder gives the decoder a single bit of information, which may prove extremely useful for the decoder strategy.
Consequently, implicit information intertwines the design of both the encoder and the decoder.
Neglecting this aspect \emph{a priori} decouples the design problem, but at the cost of a significant optimality gap (for more details, see~\cite{{santi2024remote}}). The pull-based setting also does not allow for the use of implicit information, as it restricts the available information to the knowledge of the decoder, making the problem easier to solve but incurring the same performance cost.

\subsection{Special Case: Remote Estimation}
\label{ssec:estimation}
The proposed architectures can be applied to remote estimation problems where the decoder's task is not controlling the physical process, but rather monitoring it, i.e., estimating the correct state $s_t$ at each step $t$.
This is a special case of the general \gls{decpomdp} problem, in which $\set{A}=\set{S}$ and $\mb{P}^a=\mb{P}^{a'} \, \forall  \, a, a' \in \set{A}$, and can be addressed for both the pull-based and push-based architectures. Specifically, we model this scenario by using the $0-1$ reward function: $r_{s_t,s_{t+1}}(a_t)=\delta_{s_t,a_t}$,
where $\delta_{m,n}$ is the Kronecker delta function, which is equal to $1$ if the arguments are equal and $0$ otherwise.

We observe that the results obtained for this settings hold for any reward function that depends directly on the current state and selected action, i.e., $r_{s_t,s_{t+1}}(a_t)=f(s_t,a_t)$.
Indeed, the remote estimation problem can be seen as a remote contextual bandit~\cite{pase2022rate}, and solutions to this problem reveal properties uncovered by previous work on the subject~\cite{gupta2023remote,leong2023stability}.

\section{Proposed Algorithms}\label{sec:solution}
In the following, we propose three algorithms for optimizing our \gls{cps} model and discuss the optimality properties of each solution.
We first focus on the pull-based setting, and extend the classical policy iteration approach by taking into account the \emph{a priori} belief of the decoder.
We propose two algorithms for the push-based setting: the first alternately optimizes the encoding and decoding policies, while the second is based on an intricate transformation into a single-agent POMDP.

\subsection{Pull-Based Modified Policy Iteration Scheme}
\label{sub:pull_proofs}
To solve the \gls{acnomdp} optimally,
we design an algorithm, called \gls{mpi}, which is a modified version of policy iteration considering the belief of the decoder over the state space. Note that, by Theorem~\ref{th:belief}, the knowledge of the decoder is entirely described by the tuple $\langle \Delta_t,s_{t-\Delta_t}\rangle$, which can be used as a modified state. Considering the division into phases presented in Sec.~\ref{sub:pull}, we thus define the following policies for the~decoder:
\begin{itemize}
    \item $\bm{\pi}_{\text{d}}$: the control policy, which maps each state $\langle \Delta_t,s_{t-\Delta_t}\rangle \in \{0,\ldots,T_{\max}\}\times\mc{S}$ to a control action $\pi_{\text{d}}( \Delta_t,s_{t-\Delta_t}) \in \mc{A}$;
    \item $\bm{\tau}$: the communication policy, mapping state $s_{t-\Delta_t} \in \mc{S}$ to a delay $ \tau(s_{t-\Delta_t}) \in \{0,\ldots,T_{\max}\}$ until the next update.
\end{itemize}
The above model for the communication policy $\bm{\tau}$ is more compact than a mapping from $\langle \Delta_t,s_{t-\Delta_t}\rangle$ to a binary pull decision, as the decoder will pull only after $\tau(s_{t-\Delta_t})$ steps.

\begin{figure}[t]
\vspace{-8pt}
\begin{algorithm}[H]
\caption{Pull-Based Modified Policy Iteration (MPI)}
\label{alg:pi}
\begin{algorithmic}[1]
\footnotesize

\Require $\mathbf{P},r,\beta$
\State Initialize $\mb{V}_{\text{d}}( \Delta_t,s_{t-\Delta_t}) \gets 0$, randomize $\bm{\pi}_{\text{d}}( \Delta_t,s_{t-\Delta_t})$, $\bm{\tau}$

\While {true}
    \For{$( \Delta_t,s_{t-\Delta_t})\in \mathcal{S}\times\lbrace 0,...,T_{\max} \rbrace$}
        \State $V_{\text{d}}'( \Delta_t,s_{t-\Delta_t})\gets$Update using~\eqref{eq:control_update}
    \EndFor
    \State $\mb{V}_{\text{d}}\gets \mb{V}_{\text{d}}'$\Comment{Value update step}
    \For{$\Delta_t\in\{0,\ldots,T_{\max}\}$} \Comment{Iterative step}
    \For{$s_{t-\Delta_t}\in \mathcal{S}$}
        \State $\pi_{\text{d}}'( \Delta_t,s_{t-\Delta_t})\gets$Update using~\eqref{eq:control_improvement}
        \State $\tau'(s)\gets$Update using~\eqref{eq:comm_improvement}
    \EndFor
    \EndFor
    \If {$\bm{\pi}_{\text{d}}'=\bm{\pi}_{\text{d}}\wedge\bm{\tau}'=\bm{\tau}$}
        \State\Return $\bm{\pi}_{\text{d}}, \bm{\tau}$ \Comment{Convergence}
    \Else
        \State $\bm{\pi}_{\text{d}},\bm{\tau}\gets\bm{\pi}_{\text{d}}',\bm{\tau}'$    \Comment{Policy improvement step}
    \EndIf
\EndWhile
\end{algorithmic}
\end{algorithm}
\vspace{-30pt}
\end{figure}

Since the state transitions until a pull decision are deterministic, i.e., the decoder's state always goes from $\langle \Delta_t,s_{t-\Delta_t}\rangle$ to $\langle \Delta_t+1,s_{t-\Delta_t}\rangle$ unless it requests an update, the action vector $\mb{a}_{t-\Delta_t:t-1}(s_{t-\Delta_t}|\bm{\pi}_{\text{d}})\in\mc{A}^{\Delta_t}$ can be determined in advance.
We can then compute the belief $\bm{\omega}^{\mb{a}_{t-\Delta_t:t-1}(s_{t-\Delta_t}|\bm{\pi}_{\text{d}})}_{\langle \Delta_t,s_{t-\Delta_t}\rangle}$ over the state space using~\eqref{eq:belief_full}, and the value function $V_{\text{d}}( \Delta_t,s_{t-\Delta_t})$ of the control policy $\bm{\pi}_{\text{d}}$ using the Bellman equation:
\begin{align}\label{eq:control_update}
&V_{\text{d}}( \Delta_t,s_{t-\Delta_t}) = \sum_{\mathclap{s_t,s_{t+1}\in\mc{S}}} \omega^{\mb{a}_{t-\Delta_t:t-1}(s_{t-\Delta_t}|\bm{\pi}_{\text{d}})}_{\langle \Delta_t,s_{t-\Delta_t}\rangle}(s_t)P^{\pi_{\text{d}}( \Delta_t,s_{t-\Delta_t})}_{s_t,s_{t+1}} \nonumber\\[-1\jot]
&\times \!\big[r_{s_t,s_{t+1}}(\pi_{\text{d}}( \Delta_t,s_{t-\Delta_t}))\!+\!\gamma\delta_{\tau(s_{t-\Delta_t}),\Delta_t+1}V_{\text{d}}( 1,s_{t+1}) \nonumber\\[1\jot]
&+\gamma\left(1\!-\!\delta_{\tau(s_{t-\Delta_t}),\Delta_t+1}\right)\left(V_{\text{d}}( \Delta_t+1,s_{t-\Delta_t})\!-\!\beta\right)\big].
\end{align}
Since the belief depends on previous actions, we start from $\Delta_t=0$ and iterate, using previously updated policy steps to compute the value function for larger values of $\Delta_t$:
\begin{align}\label{eq:control_improvement}
&\pi_{\text{d}}'( \Delta_t,s_{t-\Delta_t}) = \argmax_{a\in\mc{A}}\ \ \sum_{\mathclap{s_t,s_{t+1}\in\mc{S}}}\omega^{\mb{a}_{t-\Delta_t:t-1}(s_{t-\Delta_t}|\bm{\pi}'_{\text{d}})}_{\langle \Delta_t,s_{t-\Delta_t}\rangle}(s_t) \nonumber\\[-.75\jot]
&\quad \times P^a_{s_t,s_{t+1}}\bigg[ r(s_t,a,s_{t+1})+\gamma\delta_{\tau(s_{t-\Delta_t}),\Delta_t+1}V_{\text{d}}( 0,s_{t+1}) \nonumber\\[-1\jot]
&\quad +\gamma\left(1-\delta_{\tau(s_{t-\Delta_t}),\Delta_t+1}\right)\left(V_{\text{d}}( \Delta_t+1,s_{t-\Delta_t})-\beta\right)\bigg].
\end{align}
To perform the communication policy improvement step, we modified the standard update rule to consider the evolution of the Markov chain until the state is sampled again.
Given a state sequence $\mb{s}_{t+1:t+m}\in\mc{S}^{m}$, after observing state $s_{t-\Delta_t}$, the belief $\omega^{\bm{\pi}_{\text{d}}}_{\langle \Delta_t, s_{t-\Delta_t}\rangle}(\mb{s}_{t+1:t+m})$ follows from~\eqref{eq:belief_full}: 
\begin{equation}\label{eq:belief_policy}
\begin{aligned}
 \omega&^{\bm{\pi}_{\text{d}}}_{\langle \Delta_t,s_{t-\Delta_t}\rangle}(\mb{s}_{t+1:t+m})=\omega^{\mb{a}_{t-\Delta_t:t}(s_{t-\Delta_t}|\bm{\pi}_{\text{d}})}_{\langle \Delta_t,s_{t-\Delta_t}\rangle}(s_{t+1})\\
 &\times\prod_{\ell=1}^{\mathclap{m-1}}(1-\delta_{\tau(s_{t-\Delta_t}),\Delta_t+\ell})P^{\pi_{\text{d}}(\Delta_t+\ell,s_{t-\Delta_t})}_{s_{t+\ell},s_{t+\ell+1}}.
 \end{aligned}
\end{equation}
Using the whole sequence of intermediate steps in the update equations is essential to take into account that each state $\langle \Delta_t,s_{t-\Delta_t}\rangle,\,\forall \, \Delta_t>0$ is non-Markovian, as the belief distribution depends on the action sequence taken since the last communication.
Hence, the communication policy can be updated as:
\begin{align}\label{eq:comm_improvement}
&\tau'(s_t)\!=\!\argmax_{\mathclap{m\in\{0,\ldots,T_{\max}\}}}\quad\quad\ \ \sum_{\mathclap{\mb{s}_{t+1:t+m}\in\mc{S}^{m}}}\omega^{\bm{\pi}_{\text{d}}}_{\langle 0,s_t\rangle}(\mb{s}_{t+1:t+m})\bigg[\gamma^m V_{\text{d}}(0,s_{t+m})\nonumber\\
&-\gamma^m\beta +r_{s_t,s_{t+1}}(\pi_{\text{d}}(0,s_t))\!+\!\sum_{\ell=1}^{\mathclap{m-1}} \gamma^{\ell}r_{s_{t+\ell},s_{t+\ell+1}}(\pi_{\text{d}}(\ell,s_t))\bigg].
\end{align}

Note that \gls{mpi} provides an optimal solution for the pull-based setting, which however represents a restricted version of the general \gls{decpomdp} problem presented in this paper, as we will discuss in Sec.~\ref{sec:theory}.
The pseudocode for \gls{mpi} is given in Alg.~\ref{alg:pi}.

\begin{figure}
\vspace{-8pt}
\begin{algorithm}[H]
\caption{Push-Based Alternating Policy Iteration (API)}
\label{alg:api}
\begin{algorithmic}[1]
\footnotesize
\Require $\mathbf{P},r,\beta, \bm{\pi}_{\text{e}}^{(0)}$
\State Initialize $\bm{\pi}_{\text{e}}\gets\bm{\pi}_{\text{e}}^{(0)}$, randomize $\bm{\pi}_{\text{d}}$
\While {true}
    \State $\bm{\pi}_{\text{d}}'\gets$\Call{ControlPolicyIteration}{$\mathbf{P},r,\beta, \bm{\pi}_{\text{e}}$}
    \State $\bm{\pi}_{\text{e}}'\gets$\Call{CommunicationPolicyIteration}{$\mathbf{P},r,\beta, \bm{\pi}_{\text{d}}'$}
    \If {$\bm{\pi}_{\text{d}}'=\bm{\pi}_{\text{d}}\wedge\bm{\pi}_{\text{e}}'=\bm{\pi}_{\text{e}}$}
        \State\Return $\bm{\pi}_{\text{d}},\bm{\pi}_{\text{e}}$ \Comment{Convergence}
    \Else
        \State $\bm{\pi}_{\text{d}},\bm{\pi}_{\text{e}}\gets\bm{\pi}'_{\text{d}},\bm{\pi}'_{\text{e}}$ \Comment{Best response}
    \EndIf
\EndWhile
\end{algorithmic}
\end{algorithm}
\vspace{-25pt}
\end{figure}

\subsection{Push-Based Alternating Policy Iteration Scheme}\label{subsub:api}
For the push-based setting, we first design a computationally light iterative algorithm, called \gls{api}.
In this case, the training is organized into subsequent rounds, each split into two phases.
During the first phase of each round, the encoding policy is fixed, and the decoding policy is optimized, while during the second phase, the decoding policy is fixed and the encoding policy is optimized.
The procedure is repeated until both policies converge, i.e., each agent follows an optimal policy with respect to the other one.

Note that, during each phase, one agent's policy is static, so that the other agent experiences a Markovian environment and standard policy iteration can be applied.
In particular, the decoder computes the full belief in~\eqref{eq:belief_full} to obtain the expected reward for each tuple $\langle \Delta_t,s_{t-\Delta_t}\rangle$, while the problem on the encoder side is a fully observable \gls{mdp}.
As we will show in Sec.~\ref{sec:theory}, this algorithm always converges to a \gls{ne} in a finite number of steps.
The pseudocode for \gls{api} is given~in~Alg.~\ref{alg:api}.


\subsection{Push-Based Joint Policy Optimization Scheme} \label{subsub:pomdp}
We now design a more complex algorithm, called \gls{jpo}.
In this case, we transform the original problem into a single-agent POMDP, where the decision maker has full knowledge of the decoder's status and its decisions represent the actions taken by the encoder and the decoder in succession. This POMDP can be solved numerically and near-optimally using a point-based POMDP solver, such as \gls{sarsop} \cite{kurniawati_sarsop_nodate}, which is used in this paper. We are then able to exploit the piecewise linear convex structure of the value function, which allows us to have an anytime bound on the optimality of the result.
As we will show in Sec.~\ref{sec:theory}, this algorithm finds an optimal solution for the general \gls{decpomdp} problem presented in this paper, but requires an exponentially increasing computation time.
The pseudocode for \gls{jpo} is given~in~Alg.~\ref{alg:POMDP}.

\begin{figure}
\vspace{-8pt}
\begin{algorithm}[H]
\caption{Push-Based Joint Policy Optimization (JPO)}
\label{alg:POMDP}
\begin{algorithmic}[1]
\footnotesize
\Require $\langle\set{S}, \set{O}, \hat{\set{A}}, \hat{\mb{P}}, \hat{\bm{O}}, \hat{\bm{r}},  \gamma\rangle, \Omega_0, \epsilon$ \Comment{$\Omega_0$ is the set of possible initial decoder beliefs, given the initial state distribution}
\State Initialize $\Gamma$, $\underline{V}$ using Blind Lower Bound \cite{Hausrecht_incremental-methods_1997} \Comment{$\Gamma$ is the $\alpha$-vector representation of the lower bound}
\State Initialize $\overline{V}$ using Fast Informed Bound \cite{Hausrecht_value-function-approximations_2000}
\While {$\underline{V}(\omega_0) + \varepsilon < \overline{V}(\omega_0), \;for \;any\; \omega_0 \in \Omega$} 
\State Insert $w_0 \in \Omega_0$, s.t. $\underline{V}(\omega_0) + \varepsilon < \overline{V}(\omega_0)$ as the root of tree $T_{\set{R}}$.

\State \Call{Sample}{$T_{\set{R}}, \Gamma$}
\State For a subset of beliefs $\omega$ from $T_{\set{R}}$, \State \Call{Backup}{$T_{\set{R}},\Gamma,\omega$}
\State \Call{Prune}{$T_{\set{R}},\Gamma$}
\EndWhile
\State $\bm{\pi}_{\text{e},0}\gets$ Update using~\eqref{eq:first_step_joint}
\end{algorithmic}
\end{algorithm}

\vspace{-25pt}
\end{figure}

More specifically, we define the single-agent POMDP $\hat{\set{M}} = \langle\set{S}, \set{O}, \hat{\set{A}}, \hat{\mb{P}}, \hat{\bm{O}}, \hat{\bm{r}},  \gamma\rangle$, where $\set{S}$ is the same state space as in the underlying \gls{mdp}; $\set{O} = \set{S} \cup \{\chi\}$ is the observation set; $\hat{\set{A}} = \set{A} \times \set{C}^{|\set{S}|}$ is a modified action space, where each action is represented by a tuple $\langle a, \bm{c}\rangle$, where $a \in \set{A}$ represents the decoder's action and $\bm{c}$ is a vector with $|\set{S}|$ elements, each of which represents the encoder's action for each possible state; $\hat{P} = \{ \hat{P}^{\hat{a}} | \hat{a} \in \hat{\set{A}} \}$ is the set of transition matrices $\hat{P}^{\hat{a}}_{s,s'}=\Pr(s_{t+1}=s'|s_t=s, \hat{a}_t = \hat{a} = \langle a,\bm{c}  \rangle)=P^a_{s_t,s_{t+1}}$; $\hat{\bm{r}}: \set{S} \times \hat{\set{A}} \mapsto \mathbb{R}$ is the reward function defined as
\begin{equation}
\hat{r}(s,\langle a,\bm{c} \rangle)=\sum_{s'\in\set{S}}P^{a}_{s,s'}\left(r(s,a,s')-\gamma \beta \bm{c}_s \right);
\end{equation}
and $\hat{o}:\set{S}\times\set{C}^{|\set{S}|}\mapsto\set{O}$ is the observation function mapping  communication actions and states to decoder observations
\begin{align}
\hat{o}_t=\hat{o}(s_t,\bm{c}_t)= \begin{cases}
    \chi, & c_{t,s_t}=0, \\
    s_t, & c_{t,s_t}=1.
\end{cases}
\end{align}
Note that we can consider the optimal policy as a function of decoder's belief, instead of the observation history. The algorithm then identifies the optimal policy as
\begin{align}
\label{eq:POMDP_problem_joint}
\hat{\bm{\pi}}^*_{\text{joint}}=\argmax_{\hat{\bm{\pi}}:\Omega\mapsto\hat{A}} \sum_{t=0}^{\infty} \gamma^t \mathbb{E}_{\bm{\omega}_0, \hat{\bm{\pi}}}\left[\hat{r}(s_t,\hat{\bm{\pi}}(\bm{\omega}_t))\right],
\end{align}
where $\Omega$ is the probability simplex over $\set{S}$, i.e., the set of possible belief distributions, and  $\bm{\omega}_0$ is the decoder's initial belief. The joint policy can then be split into policies $\bm{\pi}_{\text{d},t}^*(\bm{\omega}_t)=a_t$ and $\bm{\pi}_{\text{e},t}^*(\bm{\omega}_{t-1},s_t)=c_t$, which represent deterministic mappings to actions that can be implemented in a distributed fashion. We can represent the optimal encoding policy as a function of the current state, the last transmitted state, and number of time steps since the last transmission, and the optimal control policy as function of the last two. In addition, the encoder policy in the first step, $\bm{\pi}_{\text{e},0}: \set{S} \mapsto \set{C}$, is computed separately as
\begin{equation}\label{eq:first_step_joint}
    {\bm\pi}_{\text{e},0}^*=\argmax_{\bm{\pi}_0:\mc{S}\mapsto\mc{C}} \sum_{{s_0\in\mc{S}}}P_0(s)\left[\underline{V}(\omega_0(\mb{P}_0,\pi_0(s_0)))-\beta\pi_0(s_0)\right],
\end{equation}
where $\mb{P}_0$ is the initial state distribution and $\underline{V}$ is the lower bound of the value function given after using the numerical algorithm to solve~\eqref{eq:POMDP_problem_joint}.

\section{Theoretical Results and Computational Limits}\label{sec:theory}
In this section, we discuss the advantages and drawbacks of the proposed algorithms from a theoretical point of view. Specifically, we first compare the pull-based and push-based architectures, showing that the optimal push-based solution is always better than the optimal pull-based one. Then, we analyze the \gls{mpi} and \gls{api} schemes. Finally, we consider the \gls{jpo} scheme, proving that it can reach an $\varepsilon$-optimal solution, although with a higher computational cost.

\subsection{Optimality Analysis}
Let $\bm{\pi}=\langle \bm{\tau},\bm{\pi}_{\text{d}}\rangle$ be a joint policy in the pull-based setting and $J^{\bm{\pi}}(s)=\mathbb{E}\left[\sum_{t=0}^\infty \gamma^tr_t \mid s_0=s,\bm{\pi}\right]$ be the long-term reward (without considering the communication cost) starting from state $s$.
We can give a compact definition of the performance of the joint policy $\bm{\pi}=\langle \bm{\tau},\bm{\pi}_{\text{d}}\rangle$:
\begin{equation}
\label{eq:tot_r}
R_{\beta}^{\bm{\pi}}=\!\sum_{\mathclap{s \in \mc{S}}} \xi(s)\!\bigg(\! J^{\bm{\pi}}(s)-\E{\sum_{n=0}^\infty\beta\gamma^{n\tau(s')}P(s'|s,\bm{\tau})\!}\bigg)\!,
\end{equation}
where $\xi(s)$ is the initial state distribution.
As a benchmark approach, we consider a periodic policy, in which the encoder communicates at regular intervals.
This solution, which is common in \glspl{cps}, sets a fixed update period $\tau \in \{0,\ldots, T_{\max}\}$, which is the same for each state in the system:
\begin{equation}
    \bm{\pi}_{\text{per}}^*(\beta) = \argmax_{\substack{\tau\in \{0,\ldots,T_{\max}\},\\\bm{\pi}_{\text{d}}:\{0,\ldots,T_{\max}\}\times\mc{S}\mapsto\mc{A}}} R_{\beta}^{\tau,\bm{\pi}_{\text{d}}}.
\end{equation}
To improve the performance, we can consider an adaptive pull-based policy, where each communicated message is associated with a different inter-transmission period, with a strategy $\bm{\tau} \in \{0,\ldots, T_{\max}\}^{|\mc{S}|}$.
The optimization problem hence becomes
\begin{equation}
\label{eq:pull_problem}
    \bm{\pi}_{\text{pull}}^*(\beta) = \argmax_{\substack{\bm{\tau}\in \{0,\ldots,T_{\max}\}^{|\mc{S}|},\\\bm{\pi}_{\text{d}}:\{0,\ldots,T_{\max}\}\times\mc{S}\mapsto\mc{A}}} R_{\beta}^{\bm{\tau},\bm{\pi}_{\text{d}}}.
\end{equation}
To solve \eqref{eq:pull_problem}, we can exploit the policy iteration strategy presented in Sec.~\ref{sub:pull}, which can reach the optimal solution is polynomial time, as expressed by Lemma~\ref{th:api_bound}.

To further improve the performance we can consider the push-based architecture,
where the communication policy $\pi_{\text{e}}( s_t,\Delta_t,s_{t-\Delta_t})$ depends on the current state $s_t$ as well as the last state update $s_{t-\Delta_t}$ and the elapsed time $\Delta_t$.
The optimal communication policy maximizes the value function
\begin{align}\label{eq:value_enc}
&V_{\text{e}}( s_t,\Delta_t,s_{t-\Delta_t})=\pi_{\text{e}}( s_t,\Delta_t,s_{t-k})(V_{\text{e}}( s_t,0,s_t)-\beta) \nonumber\\[1\jot]
&+(1-\pi_{\text{e}}( s_t,\Delta_t,s_{t-\Delta_t}))\sum_{\mathclap{s_{t+1}\in\mc{S}}}\gamma P^{\pi_{\text{d}}( \Delta_t,s_{t-\Delta_t})}_{s_ts_{t+1}} \nonumber\\[-1\jot]
&\times\!\left[\frac{r_{s_t,s_{t+1}}(\pi_{\text{d}}(\Delta_t,s_{t-\Delta_t}))}{\gamma}+V_{\text{e}}( s_{t+1},\Delta_t\!+\!1,s_{t-\Delta_t})\!\right]\!.
\end{align}
In particular, the superiority of the push-based approach over the pull-based one is proved by Theorem~\ref{th:aoi_pull}, where we write $\bm{\pi}\succeq_{\beta}\bm{\pi}'$ to denote that the joint policy $\bm{\pi}$ outperforms the joint policy $\bm{\pi}'$, i.e., that $R_{\beta}^{\bm{\pi}}\geq R_{\beta}^{\bm{\pi}'}$.

\begin{lemma}
\label{th:api_bound}
The \gls{mpi} scheme in Alg.~\ref{alg:pi} returns the optimal pull-based joint policy in polynomial time over $|\mc{A}|$ and $|\mc{S}|$.
\end{lemma}
\begin{IEEEproof}
The modified value function in~\eqref{eq:control_update} returns the pull-based policy's state value when considering the decoder's belief. Applying PI over the problem then preserves its optimality properties~\cite{howard1960dynamic}. The proof that PI is strongly polynomial was first given in~\cite[Th.~4.2]{ye2011simplex}. Each iteration, even with the modified value and update function, requires $|\mc{A}||\mc{S}|^2$ steps, resulting in a complexity increase by a factor $|\mc{S}|$, which preserves the strongly polynomial nature.
\end{IEEEproof}

\begin{theorem}
\label{th:aoi_pull}
The optimal joint policy in the push-based setting always outperforms the optimal joint policy in the pull-based setting, which outperforms any periodic policy:
\begin{equation}
\bm{\pi}^*_{\text{push}}(\beta)\succeq_{\beta}\bm{\pi}^*_{\text{pull}}(\beta)\succeq_{\beta}\bm{\pi}^*_{\text{per}}(\beta),\ \forall \ \langle \set{S}, \set{A}, \mb{P}, r, \gamma,\beta\rangle.
\end{equation}
\end{theorem}
\begin{IEEEproof}
We can first prove that the optimal pull-based policy outperforms any periodic policy by \emph{reductio ad absurdum}: we consider a hypothetical optimal interval $\bm{\pi}^*_{\text{per}}(\beta)$, which performs better than the pull-based policy for a given value of $\beta$. In this case, $\bm{\pi}^*_{\text{pull}}(\beta)$ cannot be optimal, as the vector in which all elements are equal to $\bm{\pi}^*_{\text{per}}(\beta)$ is one of the possible choices for pull-based. Then, we prove that the optimal push-based policy outperforms any pull-based policy in the same way: as the encoder can act with full knowledge of the state, the pull-based optimal policy is a possible solution to the push-based problem, and often better push-based policies exist due to an extended information structure. 
\end{IEEEproof}

\subsection{Computational and Performance Limitations}
The \gls{api} scheme presented in Sec.~\ref{subsub:api} tries to estimate the optimal value function $V_{\text{e}}$ given in \eqref{eq:value_enc} by modeling the \gls{cps} system as a Markov potential game~\cite{wang2002reinforcement}.
In particular, the encoder and the decoder act as players in a game, where the moves are the possible policies, and the payoff for each player is the expected reward in the initial state.

\begin{lemma}\label{th:nash}
The \gls{api} scheme leads to an $\varepsilon$-\gls{ne} policy in the push-based problem in polynomial time over $|\mc{A}|$ and $|\mc{S}|$.
\end{lemma}
\begin{IEEEproof}
Firstly, we can trivially prove that the considered Markov game is an exact potential game~\cite{monderer1996potential}: as the reward for the two agents is the same, the expected long-term reward is a potential function for the game. We then consider the \gls{api} scheme: each round of the iterated algorithm leads to the optimal policy when the policy of the other agent is given, due to the optimality of standard \gls{pi}. The \gls{api} scheme is then an \gls{ibr} scheme for the game, which leads to an \gls{ne} in a finite number of steps in all finite potential games due to the finite improvement path property~\cite{kukushkin2004best}.
Unfortunately, reaching an \gls{ne} in an exact potential game may require an exponential number of rounds~\cite{schaffer1991simple}, as exact potential games belong to the \gls{pls} class\footnote{The \gls{pls} class includes local optimization problem for which the cost of a solution can be computed in polynomial time and its neighborhood can be searched in polynomial time. Several \gls{ne} computation problems are in this complexity class.}, for which no polynomial-time algorithms are currently known~\cite{yannakakis2009equilibria}. However, \gls{ibr} has been shown to converge to an $\varepsilon$-\gls{ne} in $O\left(\frac{1}{\varepsilon}\right)$ iterations~\cite{christodoulou2012convergence}. As each iteration requires $|\mc{A}||\mc{S}|^2$ steps, the \gls{api} scheme is still strongly polynomial, as long as we accept approximate convergence.
\end{IEEEproof}

\begin{lemma}\label{th:push_pull}
The solution obtained by the \gls{api} scheme, $\bm{\pi}^{\text{API}}_{\text{push}}(\beta|\bm{\pi}_{\text{e}}^0)$ does not always outperform the periodic policy:
\begin{equation}
\exists\langle \set{S}, \set{A}, \mb{P}, r, \gamma,\beta \rangle,\bm{\pi}_{\text{e}}^0: \bm{\pi}^{\text{API}}_{\text{push}}(\beta|\bm{\pi}_{\text{e}}^0)\nsucceq_{\beta}\bm{\pi}^*_{\text{per}}(\beta).
\end{equation}
\end{lemma}
\begin{IEEEproof}
We can prove the theorem by considering a simple counterexample, represented by an \gls{mdp} with $5$ states and $2$ actions, whose evolution is depicted in Fig.~\ref{fig:markov_model}.\begin{figure}
    \centering
\vspace{-0.4cm}
\begin{tikzpicture}[->, shorten >=2pt, line width=0.5 pt, node distance =1 cm]
                        ]
\node (n0)  [circle,draw] {$0$};
\node (n4)  [circle,draw,below right=of n0] {$4$};
\node (n1)  [circle,draw,above right=of n0] {$1$};
\node (n2)  [circle,draw, right=of n1] {$2$};
\node (n3)  [circle,draw, right=of n2] {$3$};

\path (n0) edge [bend left] node [right]{$a_2,p=0.5$} (n4);
\path (n0) edge [bend right] node [right]{$a_2,p=0.5$} (n1);
\path (n0) edge [bend left] node [left]{$a_1$} (n1);
\path (n1) edge node [above]{$a_1$} (n2);
\path (n2) edge node [above]{$a_1$} (n3);
\path (n3) edge [bend right=110] node [below,near start]{$a_1$} (n0);
\path (n4) edge [loop right] node [right]{$a_1$} (n4);
\path (n4) edge [bend left] node [left]{$a_2$} (n0);

\path (n2) edge [loop below] node [below right]{$a_2$} (n2);
\path (n3) edge [loop below] node [below right]{$a_2$} (n3);
\path (n1) edge [loop above] node [right]{$a_2$} (n1);

\end{tikzpicture}
    \caption{A Markov model with $5$ states and $2$ actions in which the \gls{api} scheme may reach suboptimal solutions.}\vspace{-0.4cm}
    \label{fig:markov_model}
\end{figure}
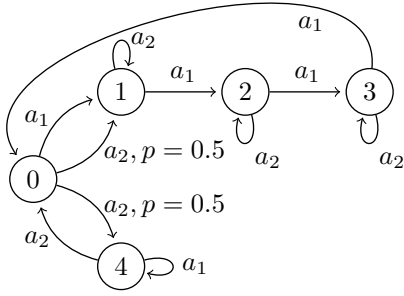
The two transition matrices corresponding to $a_1$ and $a_2$ are:
\begin{equation}
\mb{P}^{a_1} \!=\! {
\footnotesize\begin{bmatrix}
        0 & 1 & 0 & 0 & 0 \\
        0 & 0 & 1 & 0 & 0 \\
        0 & 0 & 0 & 1 & 0 \\
        1 & 0 & 0 & 0 & 0 \\
        0 & 0 & 0 & 0 & 1
    \end{bmatrix}},
\mb{P}^{a_2} \!=\!{
\footnotesize\begin{bmatrix}
        0 & 0.5 & 0 & 0 & 0.5 \\
        0 & 1 & 0 & 0 & 0 \\
        0 & 0 & 1 & 0 & 0 \\
        0 & 0 & 0 & 1 & 0 \\
        1 & 0 & 0 & 0 & 0
    \end{bmatrix}}.
\end{equation}
The reward is then always 0, except for when the environment transits to state 0, i.e., we have $r_{s,0}^a=1$ and $r_{s,s'}^a=0\ \forall s'\neq0$.

We can easily see that taking action $a_1$ in state $0$ leads to a loop with $4$ states, while taking action $a_2$ may lead to a shorter path back to the reward-giving state.
We consider an encoder policy $\tau_{\text{per}}$ that transmits in each odd step, ensuring that the decoder always knows if it lands in state $1$ or state $4$ after taking action $a_2$.
Its expected long-term reward is
\begin{equation}
 R_{\beta}^{\bm{\pi}_{\text{per}}}=\frac{2-(2\gamma+\gamma^3)\beta}{2(1-\gamma^2-\gamma^4)}.
 \end{equation}

Considering a push-based approach and applying the \gls{api} scheme, the final results depend on the initial policy of the encoder.
If the process starts from a policy that communicates often, e.g., one that always communicates the state, the algorithm will converge to the optimal joint policy, which only communicates if it deviates from the short cycle (i.e., if the system ends up in state 1).
The expected reward is
\begin{align}
R_{\beta}^{\bm{\pi}^{\text{API}}_{\text{push}}(\beta|\bm{\pi}_{\text{e}}^1)} 
&= \frac{2-\gamma\beta}{2(1-\gamma^2-\gamma^4)},
\end{align}
which is better than the periodic policy for any value of $\beta\in\mathbb{R}^+$ and $\gamma\in(0,1)$.
On the other hand, if the \gls{api} scheme starts from a policy that \emph{never} communicates, the decoder will take the conservative choice, and always take action $a_1$.
This is another \gls{ne}, as the encoder should never communicate if the decoder's policy is independent of the state.
The expected reward is
\begin{equation}
 R_{\beta}^{\bm{\pi}^{\text{API}}_{\text{push}}(\beta|\bm{\pi}_{\text{e}}^0)}=(1-\gamma^4)^{-1}.
\end{equation}
In this case, the solution provided by the \gls{api} scheme is not Pareto optimal, as it performs worse than a simple periodic policy if
\begin{equation}
 \beta<2(2+\gamma^2)^{-1}(1-\gamma^4)^{-1}.
\end{equation}
We can also easily construct a parallel \gls{mdp} for which starting from always communicating is a suboptimal solution.
\end{IEEEproof}

We now prove that the \gls{jpo} algorithm presented in Sec.~\ref{subsub:pomdp} reaches an $\varepsilon$-optimal solution for the original push-based pragmatic communication problem $\set{M}$.
\begin{theorem}\label{theorem:}
The push-based pragmatic communication problem $\set{M}$ in~\eqref{eq:unconstrained_problem}, is equivalent to the single-agent POMDP $\hat{\set{M}}$, whose optimal policy is given by~\eqref{eq:POMDP_problem_joint}.\end{theorem}

\begin{IEEEproof}Theorem~\ref{th:belief} shows that the optimal encoder policy at time $t$ can be written in the form $\bm{\pi}_{\text{e},t} \in \set{S} \times \set{O}^{t} \mapsto \set{C}$. Similarly, the information available to the decoder is $\langle o_{0:t}, a_{0,t-1} \rangle $, the history of messages and the decoder actions, and the optimal decoder policy at time $t$ is $\bm{\pi}_{\text{d},t} \in \set{O}^{t+1} \mapsto \set{A}$, as we use deterministic policies and the action history can be reconstructed from the observations. Note that the optimal communication action at time $t=0$ is given by~\eqref{eq:first_step_joint}. For the rest of the terms including the rewards and the communications costs, the optimization problem can be written, using the notation from POMDP $\hat{\set{M}}$ and the joint policy $\hat{\bm{\pi}}_t \in \set{O}^{t+1} \mapsto \set{C}^{|\set{S}|} \times \set{A}$, as
\begin{align}
&\max_{\bm{\pi}} \sum_{t=0}^{\infty} \gamma^t \mathbb{E}_{\bm{\omega}_0,\bm{\pi}}[r_{s_t,s_{t+1}}(a_t) - \gamma \beta c_{t+1}]= \max_{\bm{\pi}} \sum_{t=0}^{\infty} \gamma^t \nonumber\\[0.5\jot]
&\ \ \times \sum_{\mathclap{\langle s_t,o_{0:t}\rangle \in\set{S}\times\set{O}^{t+1}}} \Pr( s_t,o_{0:t}|\bm{\omega}_0,\bm{\pi}) \sum_{\mathclap{a_t \in \set{A}}}\pi_{\text{d},t}(a_t|o_{0:t})\sum_{\mathclap{s_{t+1}\in \set{S}}}P_{s_t,s_{t+1}}^{a_t} \nonumber\\ 
&\ \times\sum_{\mathclap{c_{t+1} \in \set{C}}}\pi_{\text{e},t+1}(c_{t+1}|\langle s_{t+1},o_{0:t}\rangle)[r(s_t,a_t,s_{t+1})-\gamma \beta c_{t+1}] \nonumber\\[-1\jot]
&= \max_{\hat{\bm{\pi}}}  \sum_{t=0}^{\infty} \gamma^t \mathbb{E}_{\bm{\omega}_0, \hat{\bm{\pi}}} \bigg[\sum_{\hat{a}_t \in \hat{\set{A}}} \hat{\pi}_t(\hat{a}_t|o_{0:t}) \hat{r}(s_t,\hat{a}_t) \bigg].\label{eq:regroup_as_expectation}
\end{align}
We can then see that $\hat{\set{M}}$ is equivalent to the original problem. This does not restrict the set of achievable policies, as all deterministic policies are separable. It is well-known that infinite horizon discounted POMDPs can be solved optimally by a stationary deterministic policy, which is a function of the belief. As a result, the initial belief $\bm{\omega}_0$ does not affect the optimal policy $\hat{\bm{\pi}}^*$.
\end{IEEEproof}



Finding the optimal solution requires learning a continuous function. Numerical algorithms in the literature~\cite{lim2023optimality} achieve an $\varepsilon$-optimal lower bound solution for a specified initial belief $\bm{\omega}_0$, i.e. they find a policy $\bm{\pi}$ such that $\underline{V}^{\bm{\pi}}(\bm{\omega}_0)\leq V^*(\bm{\omega}_0) \leq \underline{V}^{\bm{\pi}}(\bm{\omega}_0)+\varepsilon$. The set of all possible initial beliefs $\bm{\omega}_0$ for all initial encoder policies and initial state realizations is $\Omega_0$. Thus, ensuring that our policy is $\varepsilon$-optimal lower bounded $\forall \bm{\omega}_0\in\Omega_0$ is a sufficient condition for an $\varepsilon$-optimal final policy including the initial encoder policy, as can be trivially shown.

\begin{lemma}\label{lm:jpo_complexity}
The computational complexity of the \gls{jpo} algorithm is at least exponential over $|\set{S}|$.
\end{lemma}
\begin{IEEEproof}
The action space $\hat{\set{A}} = \set{A} \times \set{C}^{|\set{S}|}$ of the modified single-agent POMDP $\hat{\set{M}}$ grows exponentially with $|\set{S}|$. As the modified problem's state space $\Omega$ is larger than $\set{S}$, solving the modified problem in polynomial time is impossible, as it would require an $O(\log(|\set{A}|))$ solution to a linear optimization problem.
\end{IEEEproof}
\begin{theorem}\label{th:nexp}
Finding the optimal push-based policy $\bm{\pi}^*_{\text{push}}$ is not possible in polynomial time over $|\set{S}|$ and $|\set{A}|$.
\end{theorem}
\begin{IEEEproof}
The complexity of \glspl{decpomdp} has been studied in~\cite{bernstein2002complexity,goldman2004decentralized}.
We can easily show that the remote POMDP does not have independent observations~\cite[Def.~2]{goldman2004decentralized}, as the observation of the encoder depends on the decoder's action. According to~\cite[Cor.~3]{goldman2004decentralized}, a jointly fully observable \gls{decpomdp} is NEXP-complete\footnote{The NEXP class represents the set of problems that can be solved by non-deterministic Turing machines in time $\exp\left(N^{O(1)}\right)$, where $N$ is the size of the problem.} unless it has the independent observation property, along with other properties. As NEXP$\neq$P~\cite{seiferas1978separating}, there is no polynomial-time algorithm that can yield the solution to our problem. This is consistent with results in game theory that show that enumerating \glspl{ne} is a difficult problem~\cite{gilboa1989nash}.
\end{IEEEproof}

\subsection{The Remote Estimation Problem}
The remote estimation problem is conceptually simpler than the remote control problem, as
the decoder's actions do not affect the evolution of the system state. In the pull-based regime, the problem reduces to finding the policy $\bm{\tau}^*$, as the optimal decoder action is to guess the highest-probability state based on the belief function in~\eqref{eq:belief}.
This also simplifies the decoder's best response computation in the \gls{api} scheme: as the decoder's actions do not affect the evolution of the system, the optimal policy is the one that maximizes the immediate reward.
Given the binary definition of the reward function, the decoder's best response is then simply
\begin{equation}
\begin{aligned}
    \bm{\pi}^*_{\text{d}}(o_{0:t},\bm{\pi}_{\text{e}})= \argmax\left(\bm{\omega}_{\langle \Delta_t,s_{t-\Delta_t}\rangle}^{\bm{\pi}_{\text{e}}}\right).
\end{aligned}
\end{equation}
However, Theorem~\ref{th:nexp} still holds: due to the interactions between the agents' observations, i.e., to the existence of implicit information, even this special case belongs to the NEXP-complete complexity class~\cite[Lemma~3]{goldman2004decentralized}.

There are some restrictions of the problem for which a global optimum can be found in polynomial time even in the push-based setting.
One such case is the rate of communication minimization problem with the constraint of perfect estimation. This requires the decoder to be certain of the current state at any time: its belief over the underlying state should belong to the set of natural basis vectors of length $|\set{S}|$, i.e., $\bm{\omega}_t\in \{ \mb{e}_1, \mb{e}_2, \ldots, \mb{e}_{|\set{S}|} \}$.

\begin{figure*}
    \centering
    \subfloat[Reward (MPI).\label{fig:pull_r}]
    {
\begin{tikzpicture}

\definecolor{darkgray176}{RGB}{176,176,176}

\begin{axis}[
height = \sfheight,
width = \sfwidth,
colorbar horizontal,
colorbar style={at={(0,1.25)},anchor=south west,height=0.4cm},
colormap/viridis,
point meta max=1.85459235156934,
point meta min=0,
tick align=outside,
tick pos=left,
x grid style={darkgray176},
xmin=-0.5, xmax=20.5,
xlabel={$\beta$},
xtick style={color=black},
xtick={0,4,8,12,16,20},
xticklabels={0,0.4,0.8,1.2,1.6,2},
y dir=reverse,
y grid style={darkgray176},
ylabel={$d$},
ymin=-0.5, ymax=13.5,
ytick style={color=black},
ytick={1.1,4.1,7.1,10.1,13.1},
yticklabels={0.9,0.7,0.5,0.3,0.1}
]
\addplot graphics [includegraphics cmd=\pgfimage,xmin=-0.5, xmax=20.5, ymin=-0.5, ymax=13.5] {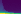};
\end{axis}

\end{tikzpicture}}\hfill
    \subfloat[Channel use (MPI). \label{fig:pull_c}]{
\begin{tikzpicture}

\definecolor{darkgray176}{RGB}{176,176,176}

\begin{axis}[
height = \sfheight,
width = \sfwidth,
colorbar horizontal,
colorbar style={at={(0,1.25)},anchor=south west,height=0.4cm},
colormap/viridis,
point meta max=1.0,
point meta min=0,
tick align=outside,
tick pos=left,
x grid style={darkgray176},
xmin=-0.5, xmax=20.5,
xlabel={$\beta$},
xtick style={color=black},
xtick={0,4,8,12,16,20},
xticklabels={0,0.4,0.8,1.2,1.6,2},
y dir=reverse,
y grid style={darkgray176},
ylabel={$d$},
ymin=-0.5, ymax=13.5,
ytick style={color=black},
ytick={1.1,4.1,7.1,10.1,13.1},
yticklabels={0.9,0.7,0.5,0.3,0.1}
]
\addplot graphics [includegraphics cmd=\pgfimage,xmin=-0.5, xmax=20.5, ymin=-0.5, ymax=13.5] {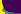};
\end{axis}

\end{tikzpicture}}\hfill
    \subfloat[PAoI (MPI, $d=0.1,\beta=1$). \label{fig:pull_peak_aoi}]{
\begin{tikzpicture}

\definecolor{darkgray176}{RGB}{176,176,176}

\begin{axis}[
height = \sfheight,
width = \sfwidth,
colorbar horizontal,
colorbar style={at={(0,1.25)},anchor=south west,height=0.4cm},
colormap/viridis,
point meta max=1,
point meta min=0,
tick align=outside,
tick pos=left,
x grid style={darkgray176},
xmin=-0.5, xmax=29.5,
xlabel=State $s$,
xtick style={color=black},
y grid style={darkgray176},
ymin=0.5, ymax=8.5,
ytick style={color=black},
ylabel={Peak AoI}
]
\addplot graphics [includegraphics cmd=\pgfimage,xmin=-0.5, xmax=29.5, ymin=-0.5, ymax=19.5] {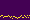};
\end{axis}
\end{tikzpicture}}\hfill
    \subfloat[Best API starting point.\label{fig:best_starting}]{
\begin{tikzpicture}

\definecolor{darkgray176}{RGB}{176,176,176}

\begin{axis}[
height = \sfheight,
width = \sfwidth,
domain=-1:1,
colorbar horizontal,
colormap={mymap}{[1pt]
rgb=(0.2667,0.0039,0.3294);
rgb=(0.1294,0.5686,0.5490);
rgb=(0.9921,0.9059,0.1451);},
colorbar sampled,
colormap access=piecewise constant,
colorbar style={samples=4,at={(0,1.25)},anchor=south west,height=0.4cm,xtick={-0.67,0,0.67},xticklabels={$\bm{\pi}_{\text{enc}}^0$,Equal,$\bm{\pi}_{\text{enc}}^1$}},
point meta max=1,
point meta min=-1,
tick align=outside,
tick pos=left,
x grid style={darkgray176},
xlabel={Beta},
xmin=-0.5, xmax=20.5,
xlabel={$\beta$},
xtick style={color=black},
xtick={0,4,8,12,16,20},
xticklabels={0,0.4,0.8,1.2,1.6,2},
y dir=reverse,
y grid style={darkgray176},
ylabel={$d$},
ymin=-0.5, ymax=13.5,
ytick style={color=black},
ytick={1.1,4.1,7.1,10.1,13.1},
yticklabels={0.9,0.7,0.5,0.3,0.1}
]
\addplot graphics [includegraphics cmd=\pgfimage,xmin=-0.5, xmax=20.5, ymin=13.5, ymax=-0.5] {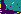};
\end{axis}

\end{tikzpicture}}\\
    \vspace{5pt}
    \centering
    \subfloat[Reward (API, best NE).\label{fig:push_r}]{
\begin{tikzpicture}

\definecolor{darkgray176}{RGB}{176,176,176}

\begin{axis}[
height = \sfheight,
width = \sfwidth,
colorbar horizontal,
colorbar style={at={(0,1.25)},anchor=south west,height=0.4cm},
colormap/viridis,
point meta max=1.85459235156934,
point meta min=0.28677148703781,
tick align=outside,
tick pos=left,
x grid style={darkgray176},
xmin=-0.5, xmax=20.5,
xlabel={$\beta$},
xtick style={color=black},
xtick={0,4,8,12,16,20},
xticklabels={0,0.4,0.8,1.2,1.6,2},
y dir=reverse,
y grid style={darkgray176},
ylabel={$d$},
ymin=-0.5, ymax=13.5,
ytick style={color=black},
ytick={1.1,4.1,7.1,10.1,13.1},
yticklabels={0.9,0.7,0.5,0.3,0.1}
]
\addplot graphics [includegraphics cmd=\pgfimage,xmin=-0.5, xmax=20.5, ymin=13.5, ymax=-0.5] {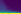};
\end{axis}

\end{tikzpicture}}\hfill
    \subfloat[Channel use (API, best NE). \label{fig:push_c}]{
\begin{tikzpicture}

\definecolor{darkgray176}{RGB}{176,176,176}

\begin{axis}[
height = \sfheight,
width = \sfwidth,
colorbar horizontal,
colorbar style={at={(0,1.25)},anchor=south west,height=0.4cm},
colormap/viridis,
point meta max=1.0,
point meta min=0.0,
tick align=outside,
tick pos=left,
x grid style={darkgray176},
xmin=-0.5, xmax=20.5,
xlabel={$\beta$},
xtick style={color=black},
xtick={0,4,8,12,16,20},
xticklabels={0,0.4,0.8,1.2,1.6,2},
y dir=reverse,
y grid style={darkgray176},
ylabel={$d$},
ymin=-0.5, ymax=13.5,
ytick style={color=black},
ytick={1.1,4.1,7.1,10.1,13.1},
yticklabels={0.9,0.7,0.5,0.3,0.1}
]
\addplot graphics [includegraphics cmd=\pgfimage,xmin=-0.5, xmax=20.5, ymin=13.5, ymax=-0.5] {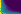};
\end{axis}

\end{tikzpicture}}\hfill
    \subfloat[PAoI (API, $d=0.1,\beta=1$). \label{fig:push_peak_aoi}]{
\begin{tikzpicture}

\definecolor{darkgray176}{RGB}{176,176,176}

\begin{axis}[
height = \sfheight,
width = \sfwidth,
colorbar horizontal,
colorbar style={at={(0,1.25)},anchor=south west,height=0.4cm},
colormap/viridis,
point meta max=1,
point meta min=0,
tick align=outside,
tick pos=left,
x grid style={darkgray176},
xlabel={State $s$},
xmin=-0.5, xmax=29.5,
xtick style={color=black},
y grid style={darkgray176},
ymin=0.5, ymax=8.5,
ytick style={color=black},
ylabel={Peak AoI}
]
\addplot graphics [includegraphics cmd=\pgfimage,xmin=-0.5, xmax=29.5, ymin=-0.5, ymax=19.5] {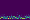};
\end{axis}
\vspace{10pt}
\end{tikzpicture}}\hfill
    \subfloat[API potential gap.\label{fig:gap}]{
\begin{tikzpicture}

\definecolor{darkgray176}{RGB}{176,176,176}

\begin{axis}[
height = \sfheight,
width = \sfwidth,
colorbar horizontal,
colorbar style={at={(0,1.25)},anchor=south west,height=0.4cm},
colormap/viridis,
point meta max=0.47452800707063,
point meta min=0,
tick align=outside,
tick pos=left,
x grid style={darkgray176},
xmin=-0.5, xmax=20.5,
xlabel={$\beta$},
xtick style={color=black},
xtick={0,4,8,12,16,20},
xticklabels={0,0.4,0.8,1.2,1.6,2},
y dir=reverse,
y grid style={darkgray176},
ylabel={$d$},
ymin=-0.5, ymax=13.5,
ytick style={color=black},
ytick={1.1,4.1,7.1,10.1,13.1},
yticklabels={0.9,0.7,0.5,0.3,0.1}
]
\addplot graphics [includegraphics cmd=\pgfimage,xmin=-0.5, xmax=20.5, ymin=13.5, ymax=-0.5] {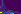};
\end{axis}

\end{tikzpicture}}
    
    \caption{Results for the remote control task in an \gls{mdp} with $30$ states.}\vspace{-0.4cm}
    \label{fig:push-pull}
\end{figure*}

\begin{theorem}
The policy that minimizes the communication cost while guaranteeing perfect state estimation at the decoder sends a message if and only if the current state realization does not correspond to the pre-communication maximum belief.
\end{theorem}
\begin{IEEEproof}
When a message is transmitted, the constraint is always met. On the other hand, the decoder may still be certain of the state even without a transmission if it uses implicit information, i.e., its knowledge of the communication policy, to rule out the states that would have resulted in a transmission according to~\eqref{eq:belief_full}. If the decoder belief at time $t-1$ is $\bm{\omega}_{t-1}=\mb{e}^{s_{t-1}}$, the pre-transmission belief at time $t$ is $\bm{\omega}'_{t}=\bm{P}\T \mb{e}^{s_{t-1}}$, while the post-transmission belief follows~\eqref{eq:belief_full}. Thus, either transmitting in all states or all but one state are the only ways to obtain a belief vector that is a natural basis vector. Any of these choices results in perfect estimation and are equivalent in terms of the evolution of the system, thus the optimal choice is to minimize the immediate transmission cost, i.e., not to transmit in the likeliest state.
\end{IEEEproof}

\begin{corollary}
If $\beta<1$, there are at least $|\mc{S}|$ local optima in the push-based remote estimation problem, even after relaxing the perfect estimation constraint.
\end{corollary}
\begin{IEEEproof}
Let us consider the following solution: the encoder sends an update for all the states, except state $s$, while the decoder estimates the received state if $\Delta_t=0$ and $s$ if $\Delta_t>0$. If $\beta<1$, this solution is an \gls{ne}: the decoder has no incentive to change, as it always estimates the correct state. If the encoder switches to not transmitting in another state $s'$,  its total reward is $1-\beta>0$. Transmitting in state $s$ would mean losing the benefit of implicit information, and having a total reward $\beta$. As the two policies are mutual best responses, the solution is an \gls{ne} for any $s$, i.e., there are at least $|\mc{S}|$ \glspl{ne}.
\end{IEEEproof}

\section{Numerical Results}\label{sec:results}
In this section, we perform numerical simulations to validate the correctness of our theoretical results.
We observe that, because of the general formulation of \glspl{mdp}, providing a comprehensive performance evaluation is not trivial.
An \gls{mdp} could be obtained by randomly generating rewards and transition matrices, but results would be hard to compare and interpret.
In order to provide a meaningful comparison, we hence consider a family of \glspl{mdp} generated by varying the number of non-zero elements in the transition matrices.

Specifically, we start from a randomly generated \gls{mdp} with deterministic transitions for each state and action and then obtain different \glspl{mdp} by splitting the transition probability among neighboring states.
The resulting \glspl{mdp} have transition matrices with different densities $d$, where the density is defined as the number of non-zero elements divided by the total number of entries of the matrix. 
In the case of the deterministic transition matrix, the density is simply $|\set{S}|^{-1}$.
If the initial deterministic transition matrix $\mb{P}^a$ has a transition from $s$ to $s'$, i.e., $P_{s,s'}^a=1$, the distribution at density $d$ is: 
\begin{equation}
    P_{s,s''}^a = \frac{\frac{(5+\delta_{s',s''})m}{5}-2|s'-s''|}{\frac{(m-1)^2}{2}+\frac{6m}{5}}, \ \forall s'' \ \text{s.t.} \ |s'-s''| < \frac{d |\set{S}|}{2},
\end{equation}
where $m=d|\set{S}|$ is the number of non-zero elements, and the index distance is over the Galois field of size $|\set{S}|$.
According to this approach, the probability of transitioning to neighboring states decreases linearly with the distance from the original transition state. Higher-density \glspl{mdp} are then simply less predictable versions of the same initial model.

\begin{figure*}
    \centering
    \subfloat[Reward (MPI).\label{fig:pull_r_estimation}]{
\begin{tikzpicture}

\definecolor{darkgray176}{RGB}{176,176,176}

\begin{axis}[
height = \sfheight,
width = \sfwidth,
point meta min=0,
point meta max=1,
colorbar horizontal,
colorbar style={at={(0,1.3)},anchor=south west,height=0.4cm},
colormap/viridis,
tick align=outside,
tick pos=left,
x grid style={darkgray176},
xmin=-0.5, xmax=20.5,
xlabel={$\beta$},
xtick style={color=black},
xtick={0,4,8,12,16,20},
xticklabels={0,0.4,0.8,1.2,1.6,2},
y dir=reverse,
y grid style={darkgray176},
ylabel={$d$},
ymin=-0.5, ymax=13.5,
ytick style={color=black},
ytick={1.1,4.1,7.1,10.1,13.1},
yticklabels={0.9,0.7,0.5,0.3,0.1}
]
\addplot graphics [includegraphics cmd=\pgfimage,xmin=-0.5, xmax=20.5, ymin=-0.5, ymax=13.5] {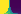};
\end{axis}

\end{tikzpicture}}\hfill
    \subfloat[Channel use (MPI). \label{fig:pull_c_estimation}]{
\begin{tikzpicture}

\definecolor{darkgray176}{RGB}{176,176,176}

\begin{axis}[
height = \sfheight,
width = \sfwidth,
point meta min=0,
point meta max=1,
colorbar horizontal,
colorbar style={at={(0,1.3)},anchor=south west,height=0.4cm},
colormap/viridis,
tick align=outside,
tick pos=left,
x grid style={darkgray176},
xmin=-0.5, xmax=20.5,
xlabel={$\beta$},
xtick style={color=black},
xtick={0,4,8,12,16,20},
xticklabels={0,0.4,0.8,1.2,1.6,2},
y dir=reverse,
y grid style={darkgray176},
ylabel={$d$},
ymin=-0.5, ymax=13.5,
ytick style={color=black},
ytick={1.1,4.1,7.1,10.1,13.1},
yticklabels={0.9,0.7,0.5,0.3,0.1}
]
\addplot graphics [includegraphics cmd=\pgfimage,xmin=-0.5, xmax=20.5, ymin=-0.5, ymax=13.5] {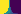};
\end{axis}

\end{tikzpicture}}\hfill
    \subfloat[PAoI (MPI, $d\!=\!0.1,\beta\!=\!1.5$).\label{fig:pull_aoi_estimation}]
    {
\begin{tikzpicture}

\definecolor{darkgray176}{RGB}{176,176,176}

\begin{axis}[
height = \sfheight,
width = \sfwidth,
colorbar horizontal,
colorbar style={at={(0,1.3)},anchor=south west,height=0.4cm},
colormap/viridis,
point meta max=1,
point meta min=0,
tick align=outside,
tick pos=left,
x grid style={darkgray176},
xlabel = States,
xmin=-0.5, xmax=29.5,
xtick style={color=black},
y grid style={darkgray176},
ymin=0.5, ymax=8.5,
ytick style={color=black},
ylabel={Peak AoI}
]
\addplot graphics [includegraphics cmd=\pgfimage,xmin=-0.5, xmax=29.5, ymin=-0.5, ymax=19.5] {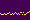};
\end{axis}

\end{tikzpicture}}\hfill
    \subfloat[Best API starting point.\label{fig:best_starting_estimation}]{
\begin{tikzpicture}

\definecolor{darkgray176}{RGB}{176,176,176}

\begin{axis}[
height = \sfheight,
width = \sfwidth,
colorbar horizontal,
colormap={mymap}{[1pt]
rgb=(0.2667,0.0039,0.3294);
rgb=(0.1294,0.5686,0.5490);
rgb=(0.9921,0.9059,0.1451);},
colorbar sampled,
colormap access=piecewise constant,
colorbar style={samples=4,at={(0,1.3)},anchor=south west,height=0.4cm,xtick={-0.67,0,0.67},xticklabels={$\bm{\pi}_{\text{enc}}^0$,Equal,$\bm{\pi}_{\text{enc}}^1$}},
point meta max=1,
point meta min=-1,
tick align=outside,
tick pos=left,
x grid style={darkgray176},
xmin=-0.5, xmax=20.5,
xlabel={$\beta$},
xtick style={color=black},
xtick={0,4,8,12,16,20},
xticklabels={0,0.4,0.8,1.2,1.6,2},
y dir=reverse,
y grid style={darkgray176},
ylabel={$d$},
ymin=-0.5, ymax=13.5,
ytick style={color=black},
ytick={1.1,4.1,7.1,10.1,13.1},
yticklabels={0.9,0.7,0.5,0.3,0.1}
]
\addplot graphics [includegraphics cmd=\pgfimage,xmin=-0.5, xmax=20.5, ymin=-0.5, ymax=13.5] {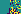};
\end{axis}

\end{tikzpicture}}\\
    \vspace{5pt}
    \centering
    \subfloat[Reward (API, best NE).\label{fig:push_r_estimation}]{
\begin{tikzpicture}

\definecolor{darkgray176}{RGB}{176,176,176}

\begin{axis}[
height = \sfheight,
width = \sfwidth,
point meta min=0,
point meta max=1,
colorbar horizontal,
colorbar style={at={(0,1.25)},anchor=south west,height=0.4cm},
colormap/viridis,
point meta max=1,
point meta min=0.05094,
tick align=outside,
tick pos=left,
x grid style={darkgray176},
xmin=-0.5, xmax=20.5,
xlabel={$\beta$},
xtick style={color=black},
xtick={0,4,8,12,16,20},
xticklabels={0,0.4,0.8,1.2,1.6,2},
y dir=reverse,
y grid style={darkgray176},
ylabel={$d$},
ymin=-0.5, ymax=13.5,
ytick style={color=black},
ytick={1.1,4.1,7.1,10.1,13.1},
yticklabels={0.9,0.7,0.5,0.3,0.1}
]
\addplot graphics [includegraphics cmd=\pgfimage,xmin=-0.5, xmax=20.5, ymin=-0.5, ymax=13.5] {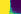};
\end{axis}

\end{tikzpicture}}\hfill
    \subfloat[Channel use (API, best NE). \label{fig:push_c_estimation}]
    {
\begin{tikzpicture}

\definecolor{darkgray176}{RGB}{176,176,176}

\begin{axis}[
height = \sfheight,
width = \sfwidth,
colorbar horizontal,
colorbar style={at={(0,1.25)},anchor=south west,height=0.4cm},
colormap/viridis,
point meta max=1,
point meta min=0,
tick align=outside,
tick pos=left,
x grid style={darkgray176},
xmin=-0.5, xmax=20.5,
xlabel={$\beta$},
xtick style={color=black},
xtick={0,4,8,12,16,20},
xticklabels={0,0.4,0.8,1.2,1.6,2},
y dir=reverse,
y grid style={darkgray176},
ylabel={$d$},
ymin=-0.5, ymax=13.5,
ytick style={color=black},
ytick={1.1,4.1,7.1,10.1,13.1},
yticklabels={0.9,0.7,0.5,0.3,0.1}
]
\addplot graphics [includegraphics cmd=\pgfimage,xmin=-0.5, xmax=20.5, ymin=-0.5, ymax=13.5] {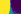};
\end{axis}

\end{tikzpicture}}\hfill
    \subfloat[PAoI (API, $d\!=\!0.1,\beta\!=\!1.5$).\label{fig:aoi_estimation}]
    {
\begin{tikzpicture}

\definecolor{darkgray176}{RGB}{176,176,176}

\begin{axis}[
height = \sfheight,
width = \sfwidth,
colorbar horizontal,
colorbar style={at={(0,1.25)},anchor=south west,height=0.4cm},
colormap/viridis,
colorbar,
colorbar style={ylabel={}},
colormap/viridis,
point meta max=1,
point meta min=0,
tick align=outside,
tick pos=left,
x grid style={darkgray176},
xlabel=States,
xmin=-0.5, xmax=29.5,
xtick style={color=black},
y grid style={darkgray176},
ymin=0.5, ymax=8.5,
ytick style={color=black},
ylabel={Peak AoI}
]
\addplot graphics [includegraphics cmd=\pgfimage,xmin=-0.5, xmax=29.5, ymin=-0.5, ymax=19.5] {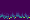};
\end{axis}

\end{tikzpicture}}\hfill
    \subfloat[API potential gap.\label{fig:gap_estimation}]{
\begin{tikzpicture}

\definecolor{darkgray176}{RGB}{176,176,176}

\begin{axis}[
height = \sfheight,
width = \sfwidth,
colorbar horizontal,
colorbar style={at={(0,1.25)},anchor=south west,height=0.4cm},
colormap/viridis,
point meta max=0.273127,
point meta min=0,
tick align=outside,
tick pos=left,
x grid style={darkgray176},
xmin=-0.5, xmax=20.5,
xlabel={$\beta$},
xtick style={color=black},
xtick={0,4,8,12,16,20},
xticklabels={0,0.4,0.8,1.2,1.6,2},
y dir=reverse,
y grid style={darkgray176},
ylabel={$d$},
ymin=-0.5, ymax=13.5,
ytick style={color=black},
ytick={1.1,4.1,7.1,10.1,13.1},
yticklabels={0.9,0.7,0.5,0.3,0.1}
]
\addplot graphics [includegraphics cmd=\pgfimage,xmin=-0.5, xmax=20.5, ymin=-0.5, ymax=13.5] {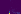};
\end{axis}

\end{tikzpicture}}

    \caption{Results for the remote estimation task in an \gls{mdp} with $30$ states.}\vspace{-0.4cm}
    \label{fig:push-pull_estimation}
\end{figure*}

In the remote control problem, we consider a two-peak model for the reward: there is a target state that gives a high reward, but also a lower-value state where the agent gets a smaller reward for reaching the state or its immediate neighborhood.
In the remote estimation special case, the reward function follows the binary model given in Sec.~\ref{ssec:estimation}.
All the code used in this work is available online\footnote{\url{https://github.com/pietro-talli/EffCom}}. 

\subsection{General Case: Remote Control}
For the remote control problem, we consider \glspl{mdp} with $30$ states, evaluating the system performance as a function of the density $d$ of the transition matrix and the communication cost $\beta$. Fig.~\ref{fig:push-pull} shows the results for the \gls{mpi} and \gls{api} schemes. In particular, we can compare the performance of the \gls{mpi} scheme in Fig.~\ref{fig:pull_r}-\subref*{fig:pull_peak_aoi} with the that of the \gls{api} scheme in Fig.~\ref{fig:best_starting}-\subref*{fig:push_peak_aoi}. 

In the pull-based setting, the system tends to transmit more often, obtaining a similar or lower reward: as expected, this is because the decoder has no additional information to decide when to request new updates other than the last transmitted state. This is also evident in the peak \gls{aoi} (PAoI), which is a deterministic function of the last observed state in the pull-based setting. While the channel use is monotonically decreasing in $\beta$, the trade-off between the reward and the channel use is more complex when we consider it as a function of the density $d$.

In the push-based setting, we considered two different starting points for the \gls{api} scheme, $\bm{\pi}_{\text{e}}^0$ and $\bm{\pi}_{\text{e}}^1$, which correspond to the two extreme policies of never and always transmitting, respectively. Fig.~\ref{fig:best_starting} shows that the two starting points often reach the same solution (within a margin of $\varepsilon=10^{-3}$) if the communication cost is high, and reach different equilibria otherwise. Fig.~\ref{fig:gap} also shows the potential gap between the two \glspl{ne}, which is generally small, except for a few cases.

\subsection{Special Case: Remote Estimation}
We perform the same analysis for the remote estimation problem. In this case, the transition matrix is fixed and does not depend on the decoder's actions.
Fig.~\ref{fig:push-pull_estimation} shows the results in a structure analogous to that of Fig.~\ref{fig:push-pull}. A significant difference with respect to the remote control problem is that, in the remote estimation case, we begin to see a trade-off between the estimation error and the channel use only at higher communication costs.

Another difference is that, at lower values of $d$, the system is able to take advantage of the more predictable transitions; and thus, can reduce the channel use even at lower values of $\beta$. This is in sharp contrast with the control case, as there is no monotonic relationship between $d$ and the channel use in that case: this is because the decoder is able to affect the state transitions, and thus the probability of future communications, by changing its policy, leading to a more complicated interaction between the control and communication policies. On the other hand, the channel use becomes a monotonically increasing function of $d$ as well as a decreasing function of $\beta$ in the estimation case, as the decoder can only estimate the state but not affect its evolution. This also leads solution reached by the \gls{api} scheme, under both starting points, to be extremely close to the solution of the pull-based \gls{mpi} scheme.

\subsection{Optimality Gap}
\begin{figure}
\begin{tikzpicture}
\begin{axis}[
    width=0.99\linewidth,
    height=0.56\linewidth,
    xlabel={Average Channel Use},
    ylabel={Average Reward},
    xmin=0, xmax=1,
    ymin=1.2, ymax=1.9,
    xtick={0.0,0.1,0.2,0.3,0.4,0.5,0.6,0.7,0.8,0.9,1.0},
    ytick={1.0,1.1,1.2,1.3,1.4,1.5,1.6,1.7,1.8,1.9,2.0},
    legend pos=south east,
    ymajorgrids=true,
    xmajorgrids=true,
    grid style=dashed
]
\addplot[color={lightgray},dashed]
table{
0   1.815534492968956
1   1.815534492968956
};
\addplot[
    color=color1,
    mark=x,
    ]
    table {figures/POMDP_control_2.dat};
\addplot[
    color=color2,
    mark=triangle,
    ]
    table {figures/control_push_always.dat};
\addplot[
    color=color3,
    mark=triangle,
    mark options={rotate=180},
    ]
    table {figures/control_push_never.dat};
\addplot[
    color=color4,
    mark=o,
    ]
    table {figures/control_pull.dat};
    \legend{Ideal, JPO, API ($\bm{\pi}_{\text{e}}^1$), API ($\bm{\pi}_{\text{e}}^0$), MPI}
\end{axis}
\end{tikzpicture}
\vspace{-0.2cm}
\caption{The trade-off curves between the average reward and the average channel use in the control task with 10 states. The density of the transition matrix is $d=0.9$.}\vspace{-0.4cm}
\label{fig:reward_vs_communication}
\end{figure}
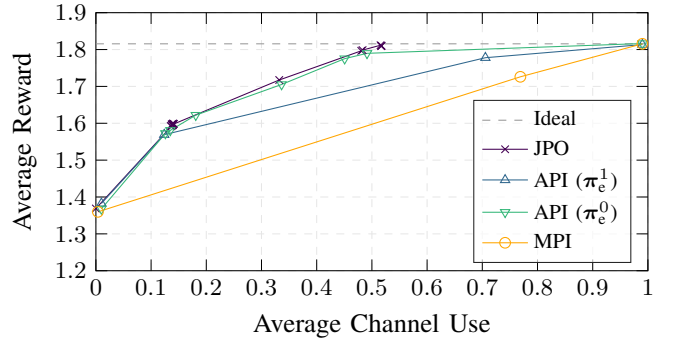

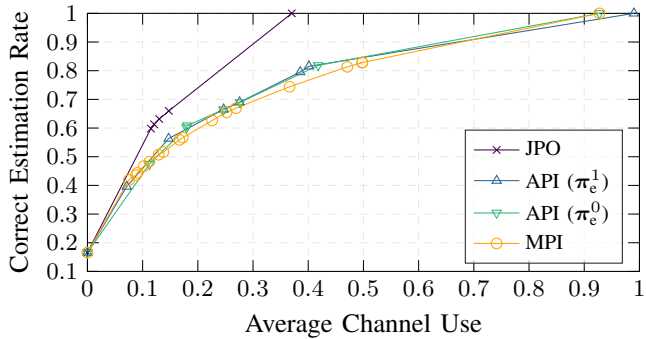
\begin{figure}
\begin{tikzpicture}
\begin{axis}[
    width=0.99\linewidth,
    height=0.56\linewidth,
    xlabel={Average Channel Use},
    ylabel={Correct Estimation Rate},
    xmin=0, xmax=1,
    ymin=0.1, ymax=1.0,
    xtick={0.0,0.1,0.2,0.3,0.4,0.5,0.6,0.7,0.8,0.9,1.0},
    ytick={0.0,0.1,0.2,0.3,0.4,0.5,0.6,0.7,0.8,0.9,1.0},
    legend pos=south east,
    ymajorgrids=true,
    xmajorgrids=true,
    grid style=dashed,
]

\addplot[
    color=color1,
    mark=x,
    ]
    table {figures/POMDP_estimation_2.dat};


\addplot[
    color=color2,
    mark=triangle,
    ]
    table {figures/api_always.dat};
\addplot[
    color=color3,
    mark=triangle,
    mark options={rotate=180}
    ]
    table {figures/api_never.dat};
    \addplot[
    color=color4,
    mark=o,
    ]
    table {figures/Pull_estimation.dat};
    \legend{JPO, API ($\bm{\pi}_{\text{e}}^1$), API ($\bm{\pi}_{\text{e}}^0$), MPI}

\end{axis}
\end{tikzpicture}
\vspace{-0.2cm}
\caption{The trade-off curves between the average correct estimation rate and the average channel use in the estimation task with 10 states. The density of the transition matrix is $d=0.19$ and the matrix was built \emph{ad hoc}.}\vspace{-0.4cm}
\label{fig:reward_vs_communication_estimation}
\end{figure}

In the following, we compare the \gls{mpi} and the \gls{api} schemes with the \gls{jpo} scheme in the two cases. Due to computational issues, we consider a single, smaller \gls{mdp}, with $|\set{S}|=10$, plotting the trade-off curves between the average reward and the channel use. Each point in Figs.~\ref{fig:reward_vs_communication} and~\ref{fig:reward_vs_communication_estimation} corresponds to a different value of $\beta$. However, a different value of $\beta$ does not necessarily result in a different solution\footnote{Deterministic policies can be combined with time-sharing, i.e., every time we reach a certain belief, we proceed with one of the policies with a certain probability and with the other with another probability. This allows us to obtain any convex combination of the performance of any two policies obtained using a specific algorithm, and thus the Pareto region is convex.}.

We compare the Pareto frontiers between the reward and the channel use for the three proposed algorithms in Fig.~\ref{fig:reward_vs_communication}. For all the algorithms, the performance is trivially similar when $\beta$ is high and the encoder almost never communicates. As the communication cost increases, the \gls{jpo} and \gls{api} schemes with $\bm{\pi}_{\text{e}}^0$ initialization achieve similar performances, showing that the \gls{api} scheme converges almost to the Pareto-efficient \gls{ne}. When it is initialized with $\bm{\pi}_{\text{e}}^1$, the \gls{api} scheme achieves a lower reward, converging to a worse \gls{ne}. However, the performance in the latter case still Pareto dominates the \gls{mpi} scheme, showing that the performance cost of restricting the problem to the pull-based case is significant. The results in Fig.~\ref{fig:reward_vs_communication} are also consistent with Fig.~\ref{fig:best_starting}, as the $\bm{\pi}_{\text{e}}^0$ initialization seems to be a better starting point for the \gls{api} scheme than $\bm{\pi}_{\text{e}}^1$ in the larger \gls{mdp}.

Conversely, for the remote estimation problem, we can see that the \gls{api} scheme is far from the optimum. Fig.~\ref{fig:reward_vs_communication_estimation} shows the \gls{jpo} scheme obtaining a very good trade-off between the reward and the channel use. However, both \gls{api} initializations have a very poor performance, and that the iterative procedure terminates in a relatively bad \gls{ne} in both cases. The working points obtained in this scenario are similar to the pull-based strategy, which we have proved to be a sub-optimal solution to the remote estimation problem. A possible explanation of this behavior is that a high level of cooperation through implicit information is more difficult to obtain, as the decoder has fewer ways to influence the trajectory across the state space. Other approaches might mitigate this issue by using different initializations, but the remote estimation problem seems to be generally harder for iterative policies.

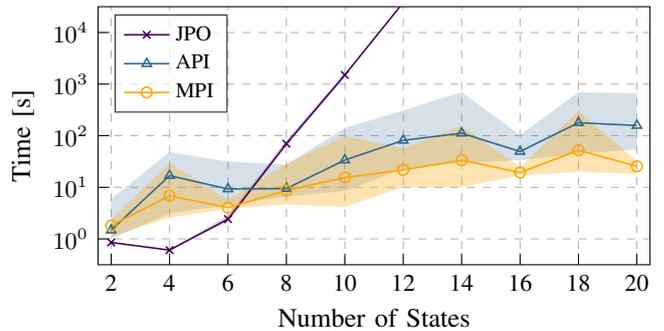
\begin{figure}[t]
    \vspace{0.19cm}
    \centering
\begin{tikzpicture}

\definecolor{darkgray176}{RGB}{176,176,176}
\definecolor{darkorange25512714}{RGB}{255,127,14}
\definecolor{steelblue31119180}{RGB}{31,119,180}

\begin{axis}[
legend pos=north west,
ymajorgrids=true,
xmajorgrids=true,
grid style=dashed,
width = 0.99\linewidth,
height=0.56\linewidth,
tick align=inside,
tick pos=left,
x grid style={darkgray176},
xmin=1.55, xmax=20.45,
xtick style={color=black},
y grid style={darkgray176},
ymin=-0.5, ymax=4.5,
xtick={2,4,6,8,10,12,14,16,18,20},
xticklabels={2,4,6,8,10,12,14,16,18,20},
ytick={0,1,2,3,4,5},
yticklabels={$10^0$,$10^1$,$10^2$,$10^3$,$10^4$,$10^5$},
ytick style={color=black},
ylabel = {Time [s]},
xlabel = Number of States
]

\path [fill=color1, fill opacity=0.2]
(axis cs:2,0.503609532)
--(axis cs:4,-0.198466991)
--(axis cs:6,0.326243566)
--(axis cs:8,1.798415884)
--(axis cs:10,3.121442805)
--(axis cs:10,3.216196251)
--(axis cs:8,1.917146823)
--(axis cs:6,0.451424455)
--(axis cs:4,-0.197032771)
--(axis cs:2,0.500876383)
--cycle;

\path [fill=color2, fill opacity=0.2]
(axis cs:2,0.004144591)
--(axis cs:4,0.492771228)
--(axis cs:6,0.656039612)
--(axis cs:8,0.823846351)
--(axis cs:10,0.926626624)
--(axis cs:12,1.359693787)
--(axis cs:14,1.473905548)
--(axis cs:16,1.523674229)
--(axis cs:18,1.647313808)
--(axis cs:20,1.735943624)
--(axis cs:20,2.822359535)
--(axis cs:18,2.841419099)
--(axis cs:16,2.013606001)
--(axis cs:14,2.844344469)
--(axis cs:12,2.48867551)
--(axis cs:10,2.150275598)
--(axis cs:8,1.440090046)
--(axis cs:6,1.49880969)
--(axis cs:4,1.684027668)
--(axis cs:2,0.777535781)
--cycle;

\path [fill=color4, fill opacity=0.3]
(axis cs:2,0.04997836)
--(axis cs:4,0.41289386)
--(axis cs:6,0.585817059)
--(axis cs:8,0.668216755)
--(axis cs:10,0.620586531)
--(axis cs:12,1.003786662)
--(axis cs:14,1.003705955)
--(axis cs:16,1.231311152)
--(axis cs:18,1.301112714)
--(axis cs:20,1.260965355)
--(axis cs:20,1.508105525)
--(axis cs:18,2.476580666)
--(axis cs:16,1.392424641)
--(axis cs:14,2.192579237)
--(axis cs:12,1.758001082)
--(axis cs:10,1.991754237)
--(axis cs:8,1.46304341)
--(axis cs:6,0.760323709)
--(axis cs:4,1.468440617)
--(axis cs:2,0.370967285)
--cycle;

\addplot [semithick, color1, mark=x]
table {%
2 -0.068355997
4 -0.21627801
6 0.384001167
8 1.845264451
10 3.17709717
12 4.57877019127
};

\addplot [semithick, color2, mark=triangle]
table {%
2 0.170159906
4 1.226702056
6 0.968021006
8 0.975433311
10 1.529362123
12 1.910360568
14 2.049226323
16 1.694052767
18 2.251850402
20 2.198115806
};

\addplot [semithick, color4, mark=o]
table {%
2 0.257618643
4 0.837278626
6 0.606371761
8 0.939528222
10 1.190078719
12 1.338279407
14 1.526433881
16 1.286915309
18 1.719056648
20 1.408095264
};
\legend{JPO, API, MPI}
\end{axis}

\end{tikzpicture}
    \caption{Running time of the proposed algorithms as a function of the size of the \gls{mdp}.}\vspace{-0.4cm}
    \label{fig:time complexity}
\end{figure}

Finally, we can confirm the asymptotic running time of the algorithms: Fig.~\ref{fig:time complexity} represents the running time, averaged over 10 random \glspl{mdp}, of the three algorithms, over the same hardware. While the implementation of the \gls{jpo} scheme is more efficient due to the use of Python libraries that exploit parallel computation, its running time grows exponentially, taking hours even for relatively small \gls{mdp}. On the other hand, the \gls{mpi} and \gls{api} schemes show a polynomial growth, corresponding to a sub-linear increase over the logarithmic plot, even though the specific implementation is relatively inefficient.

\section{Conclusions}\label{sec:conc}
In this paper, we have developed a theoretical framework that provides general results and theoretical limits for the efficiency of pragmatic communication. Our model considers the estimation and control of finite-state Markov processes over costly zero-delay communication channels, and includes two decision-making architectures, in which communication is either pull-based or push-based.
We proposed three algorithms to optimize our system, discussed different optimality solution concepts associated with these algorithms, and shed light on their computational complexity and fundamental limits. We showed that while the optimal solution can be reached in polynomial time in the pull-based case, which represents a restricted version of the problem with a significant performance gap, the push-based case poses a trade-off between global optimality and computational complexity.

We should highlight that the trade-off between communication cost and application performance becomes even more important for safety-critical, high-throughput systems such as autonomous vehicles, proving the need for pragmatic communication. The framework developed in this paper is flexible enough to allow for several future extensions. These might include more complex scenarios involving communication impairments such as time delay, packet loss, and channel noise. Another interesting extension would be to consider multiple nodes, which may have different information about the system state, requiring them to reason about each other's beliefs to coordinate their actions and transmissions over the shared channel.

\bibliographystyle{IEEEtran}
\bibliography{bibliography}

\end{document}